\documentclass[12pt]{article}

\usepackage{ytableau}
\usepackage[all]{xy}
\usepackage{tikz, tikz-cd}
\usepackage{mathdots}
\usepackage{savesym}
\usepackage{cite}
\usepackage{wasysym}
\usepackage{amscd}
\usepackage{graphicx}
\usepackage{pifont}
\usepackage{float}
\usepackage{subfig}
\usepackage{multicol}
\usepackage{amsfonts}
\usepackage{picture}
\usepackage{amssymb}
\savesymbol{iint}
\savesymbol{iiint}
\usepackage{amsmath}
\restoresymbol{TXF}{iint}
\restoresymbol{TXF}{iiint}
\usepackage{color}
\usepackage{clay}
\usepackage{hyperref}
\usepackage{slashed}
\hypersetup{colorlinks=true}
\hypersetup{linkcolor=black}
\hypersetup{citecolor=black}
\hypersetup{urlcolor=black}
\usepackage{setspace}
\usepackage{multirow}
\usepackage{wrapfig}
\usepackage{verbatim}
\numberwithin{equation}{section}

\usepackage{suffix}

\begin{document}

\institution{Fellows}{\centerline{${}^{1}$Society of Fellows, Harvard University, Cambridge, MA, USA}}
\institution{HarvardU}{\centerline{${}^{2}$Jefferson Physical Laboratory, Harvard University, Cambridge, MA, USA}}

\title{Asymptotics of Ground State Degeneracies in Quiver Quantum Mechanics}

\authors{Clay C\'{o}rdova\worksat{\Fellows}\footnote{e-mail: {\tt cordova@physics.harvard.edu}}  and Shu-Heng Shao\worksat{\HarvardU}\footnote{e-mail: {\tt shshao@physics.harvard.edu}} }

\abstract{We study the growth of the ground state degeneracy in the Kronecker model of quiver quantum mechanics.  This is the simplest quiver with two gauge groups and bifundamental matter fields, and appears universally in the context of BPS state counting in four-dimensional $\mathcal{N}=2$ systems.  For large ranks, the ground state degeneracy is exponential with slope a modular function that we are able to compute at integral values of its argument.       We also observe that the exponential of the slope is an algebraic number and determine its associated algebraic equation explicitly in several examples.  The speed of growth of the degeneracies, together with various physical features of the bound states, suggests a dual string interpretation.}

\date{March 2015}
\maketitle

\tableofcontents

\section{Introduction}

The entropy of a quantum system is a basic thermodynamic observable.  In conformal field theory in $d$ spacetime dimensions, in finite spatial volume $V$, dimensional analysis constrains the growth of the entropy $\mathcal{S}$ with energy $E$ to take the form 
\begin{equation}
\mathcal{S}\sim V^{1/d}E^{(d-1)/d}~. \label{entropy}
\end{equation}
In particular, the entropy grows slower than linearly with energy.  By contrast, in quantum field theory in infinite spatial volume, the thermodynamics is much more subtle.  Spatially large stable states, in general have a growth in energy which is faster than \eqref{entropy}, and few universal results are known (see, for example, \cite{Galakhov:2013oja}).

Motivated by these general thermodynamic considerations, in this work we study a non-relativistic supersymmetric quantum mechanics problem known as the Kronecker model.  This model occurs universally in particle counting problems in four-dimensional $\mathcal{N}=2$ field theories and supergravities where it arrises as the low-energy non-relativistic effective theory of BPS dyons or black holes  \cite{ Douglas:2000ah, Douglas:2000qw, Fiol:2000wx, Fiol:2000pd, Denef:2002ru, Denef:2007vg, Alim:2011kw, Manschot:2012rx, Cecotti:2012sf, Galakhov:2013oja, Chuang:2013wt, Cordova:2013bza}.  In this context, each ground state of the quantum mechanics is reinterpreted as a stable four-dimensional single-particle state.  The growth of the ground state degeneracy for large charges thus probes the infinite volume thermodynamics of the field theory.

\begin{figure}[here!]
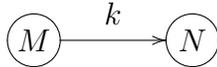

  \centering
\subfloat{
\xy  0;<1pt,0pt>:<0pt,-1pt>::
(-300,0) *+{M}*\cir<10pt>{} ="1",
(-240,0) *+{N}*\cir<10pt>{} ="2",
(-270, -10) *+{k} ="b",
\ar @{->} "1"; "2"
\endxy}
  \caption{The Kronecker quiver with $k$ arrows.  This supersymmetric quantum mechanics describes $M$ superparticles each with charge $\gamma_{1}$ and $N$ superparticles each with charge $\gamma_{2}$ with Dirac pairing $\langle \gamma_{1}, \gamma_{2}\rangle =k.$}\label{fig:kronecker}
\end{figure}

The Kronecker model of interest describes a multi-particle system composed of two distinct species of (super)particles interacting by long range electromagnetic forces.    The strength of these interactions is invariantly characterized by the integral Dirac pairing of the electromagnetic charges
\begin{equation}
\langle \gamma_{1}, \gamma_{2}\rangle =k>0~.
\end{equation}
We investigate the spectrum of $M$ particles of type one and $N$ particles of type two.  This system and its interactions are encoded in the Kronecker quiver illustrated in Figure \ref{fig:kronecker}.\footnote{An explicit expression for the Hamiltonian of this system may be found, for instance, in \cite {Denef:2002ru}.}

We focus on the ground state degeneracy of these models.  We denote this degeneracy as $\Omega(M,N,k).$  These ground states are supersymmetric and their degeneracies have been studied from a variety of perspectives, including quantum groups \cite{2003InMat.152..349R},  wall-crossing formulas \cite{Kontsevich:2008fj, Gross, Reineke}, spectral networks \cite{Galakhov:2013oja, Galakhov:2014xba}, equivariant cohomology \cite{Weist:2009, Weist:2012}, and supersymmetric localization \cite{Hwang:2014uwa, Cordova:2014oxa, Hori:2014tda, Cordova:2015qka}.

Our aim is to understand the growth in the degeneracy $\Omega(M,N,k)$ for large ranks $M$,  and $N.$  We study this limit with fixed $k$ and with fixed limiting ratio $N/M\rightarrow r.$   Known results, from the special case where $r=1,$ indicate that these degeneracies grow exponentially \cite{Weist:2009, Galakhov:2013oja, Kim:2015oxa}.   Based on this evidence, it was conjectured in \cite{Weist:2009} that there exists a \emph{slope function} $S(r,k)$ governing the asymptotics of the degeneracy at general $r,$
\begin{equation}
\lim_{M\rightarrow \infty}\frac{1}{M}\log\Big(\Omega(M+m,Mr+n,k)\Big)\equiv S(r,k)~.
\end{equation}
In particular, this function is claimed to be independent of the offset $(m,n)$ and depends only on the asymptotic ratio $r$ and number of arrows $k$ appearing in the quiver.  

As we motivate in \S \ref{sec:kron}, it is useful to express the slope function $S(r,k)$ in terms of an auxiliary function $G(r,k)$ as
\begin{equation}
S(r,k)=\Bigg(\sqrt{\frac{kr-r^{2}-1}{k-2}}\Bigg)\Bigg( (k-1)^{2}\log((k-1)^{2})-(k^{2}-2k)\log(k^{2}-2k)\Bigg)G(r,k)~. \label{gdefintro}
\end{equation}
The known exact results from the case $r=1$ are then summarized by $G(1,k)=1.$\footnote{  In \cite{weistthesis, Weist:2009} it was further conjectured that $G(r,k)=1$ for all $r$.  We find, by direct calculation, that this further conjecture is false.  }

Our main new results presented in \S \ref{sec:explicit} are explicit calculations of the slope function $S(r,k)$ (or equivalently the function $G(r,k)$) in the special case where the ratio $r$ is a general non-negative integer.  In particular in all such examples, we verify that the degeneracies indeed grow exponentially, and we find that the function $G(r,k)$ is not constant.   These calculations are possible thanks to a new formula \cite{Cordova:2015qka} which provides an explicit expression for all degeneracies of the form $\Omega(M,Mr+1,k)$ for integer $r,$ and hence enables us to explore the large rank regime of these models.  We also provide evidence that the slope is independent of the offset using wall-crossing formulas in \S \ref{sec:wall}.

The quantity $G(r,k)$ appearing in \eqref{gdefintro} is an interesting function of the ratio $r.$  As we review in \S \ref{sec:const}, dualities in the Kronecker models enable us to change $r$ without changing the ground state degeneracies.  This implies the following modular identities
\begin{equation}
G(r,k)=G(1/r,k)= G(k-1/r,k)~.
\end{equation}
These modular constraints, combined with our exact calculations at integral $r,$ indicate that the slope demonstrates intricate oscillatory behavior for large and small values of the ratio.\footnote{In \cite{weistthesis} a uniqueness theorem $G(r,k)=1$ was proven under certain continuity assumptions on $G(r,k).$  The oscillatory behavior we observe violates these continuity assumptions and hence invalidates the uniqueness theorem.  See \S \ref{sec:solve} for discussion. } See Figure \ref{fig:g} for an illustration of this behavior.

In \S \ref{sec:alg} we explore the number theoretic properties of the slope function.  We find that, for all cases that we have studied, $\exp(S(r,k))$ is an algebraic number, i.e. it solves an algebraic equation with rational coefficients.   Even for small $r$ and $k$, the resulting equations are striking in their complexity, with unexpected coefficients.  For example, when $(r,k)=(2,4)$ we find that $\exp(S(2,4))$ is the positive solution to
\begin{equation}
x^{2}-\frac{53793390359}{1088391168}x-\frac{823543}{12230590464}=0~.
\end{equation}
It would be interesting to understand a physical or geometric origin of these equations directly, perhaps by relating them to identities obeyed by generating functions of threshold bound states \cite{Kontsevich:2008fj, Gross, Reineke, Galakhov:2013oja, Galakhov:2014xba, Tom}, or to enumerative Calabi-Yau geometry.

Finally, before delving into the details, we briefly return to our motivating physical question and take stock of the properties of the ground states when they are interpreted as stable particles of four-dimensional field theories.  In that context the ranks $M$ and $N$ are linearly related to electric and magnetic charges $Q$, and hence (via BPS bounds) to particle masses $m$ (or equivalently energies $E$).  Thus, we have the scaling relations   
\begin{equation}
M\sim N \sim Q \sim m \sim E~.
\end{equation}
The general properties of the states in question are then as follows.
\begin{itemize}
\item The physical radius $R$ of the states grows linearly with the ranks $M$ and $N$ \cite{Galakhov:2013oja}, or equivalently linearly in mass $m$ 
\begin{equation}
R\sim m~.
\end{equation}
\item The particles lie on Regge trajectories \cite{Cordova:2015vma}.  In other words, the states of largest angular momentum $J$ at fixed mass $m$ obey a relation
\begin{equation}
J\sim m^{2}~.
\end{equation}
\item There is an exponential degeneracy of particle states with entropy growth linear in mass (so that \eqref{entropy} is violated)
\begin{equation}
\mathcal{S}\sim m \sim \sqrt{J}~.
\end{equation}
\end{itemize}
Taken as a whole, these features suggest the existence of a dual string model for these bound states, where the Regge behavior and exponential degeneracy are manifest.  In that context the slope function $S(r,k),$ which plays a primary role in our analysis, would then be reinterpreted in terms of the central charge of the dual world sheet string theory.  It would be satisfying to determine this string model explicitly, and we leave this as a potential avenue for future investigation.

\section{Kronecker Models and Their Indices}
\label{sec:kron}
In this section, we review the Kronecker models and their degeneracies $\Omega(M,N,k)$.  In  \S \ref{sec:conjecture} we state a conjecture concerning the behavior of these degeneracies for large ranks.  

We begin with the Kronecker quiver illustrated in Figure \ref{fig:kronecker}.  This system is a gauged $\mathcal{N}=4$ quantum mechanics.  At each node, there are vector multiplets with unitary gauge groups of ranks $M$ and $N,$ respectively.  The arrows of the quiver are bifundametal chiral multiplet matter fields.  See, for instance \cite {Denef:2002ru}, for the explicit Hamiltonian of this system. The quantity of interest, $\Omega(M,N,k),$ is the Witten index of this system.

In general, the ground states of the Kronecker model occur at threshold and are challenging to explicitly determine.  However, in the special case where $M$ and $N$ are coprime, the system is gapped and the index $\Omega(M,N,k)$ admits a simple geometric interpretation.  

To describe this correspondence, we first introduce the classical Higgs branch moduli space $\mathcal{M}^k_{M,N}$.  This moduli space is parameterized by the chiral multiplet fields $\Phi_{i}$ ($i=1,\cdots, k$) which have constant expectation values.  Thus, they specify linear maps 
\begin{equation}
\Phi_{i}: \mathbb{C}^{M}\rightarrow \mathbb{C}^{N}~.
\end{equation}
On the maps $\Phi_{i}$ we enforce the D-term equations 
\begin{equation}
\sum_{i=1}^{k}\Phi_{i}^{\dagger}\circ\Phi_{i}=\zeta I_{M}~,\hspace{.5in}\sum_{i=1}^{k}\Phi_{i}\circ\Phi_{i}^{\dagger}=\frac{M\zeta}{N} I_{N}~, \label{dterm}
\end{equation}
where $\zeta>0$ is the Fayet-Iliopoulos parameter,\footnote{When $\zeta<0$ all moduli spaces are empty, demonstrating wall-crossing.  See \S \ref{sec:wall} for discussion.} and $I_{L}$ is the $L\times L$ identity matrix.  To obtain the desired moduli space, we now quotient by the gauge group $U(M)\times U(N)$ acting on the $\Phi_{i}$ via the bifundamental representation
\begin{equation}
\mathcal{M}^k_{M,N} \equiv \left\{\Phi_{i}~\Bigg\vert \sum_{i=1}^{k}\Phi_{i}^{\dagger}\circ\Phi_{i}=\zeta I_{M}~,\hspace{.25in}\sum_{i=1}^{k}\Phi_{i}\circ\Phi_{i}^{\dagger}=\frac{M\zeta}{N} I_{N}\right\}/U(M)\times U(N)~.
\end{equation}
When $M$ and $N$ are coprime, these moduli spaces are smooth, compact, K\"{a}her manifolds. In this case, the complex dimension of the moduli space may be easily computed by subtracting the dimension of the gauge groups from the dimension of the space of chiral fields\footnote{The offset by one is due to the fact that an overall $u(1)$ in the gauge group does not act on the bifundamental chiral multiplets.}
\begin{equation}
\mathrm{dim}\left(\mathcal{M}^k_{M,N} \right)=kMN-M^{2}-N^{2}+1~. \label{dimform}
\end{equation}

As usual in supersymmetric quantum mechanics, the ground states are in one-to-one correspondence with the cohomology of the moduli space $\mathcal{M}^k_{M,N}$, and the index $\Omega(M,N,k)$ is the Euler characteristic.  In this particular case, we can say more due to a vanishing theorem constraining the Hodge decomposition of the cohomology \cite{2003InMat.152..349R}
\begin{equation}
h^{p,q}\left(\mathcal{M}^k_{M,N}\right)=0~, \hspace{.5in} \mathrm{if}~~p\neq q~. \label{vanish}
\end{equation}
The index $\Omega(M,N,k)$ is then
\begin{equation}
\Omega(M,N,k)=\chi(\mathcal{M}^k_{M,N})= \sum_{p\geq0} h^{p,p}\left(\mathcal{M}^k_{M,N}\right)~.
\end{equation}
Thus, as a consequence of the vanishing theorem \eqref{vanish}, all ground states of the model are bosons, and the index $\Omega(M,N,k)$ computes the absolute degeneracy of the ground states.

\subsection{Indices as a Function of $k$}
\label{sec:kvar}

The ground state degeneracies show significant dependence on the number of arrows $k$ in the quiver.  Qualitatively, there are three distinct cases $k=1,$ $k=2,$ and $k>2,$ with increasing $k$ demonstrating increasing complexity.  

One way to understand this phenomenon is to examine the moduli space when $M=N$.  In that case, generically, (i.e. on an open set in the moduli space) at least one of the maps $\Phi_{i}$ is invertible.  We may then remove some of the gauge redundancy by fixing one such map to the identity matrix.  After doing so, we must study $k-1$ linear maps modulo conjugation.  For $k=1$ this problem is trivial.  For $k=2,$ this problem is solved by the Jordan decomposition theorem.  For $k>2$ this is a notoriously wild representation theory problem with no known exact solution.

\begin{figure}[h]
\begin{center}
\subfloat[$k=1$]{
\includegraphics[width=.3\textwidth]{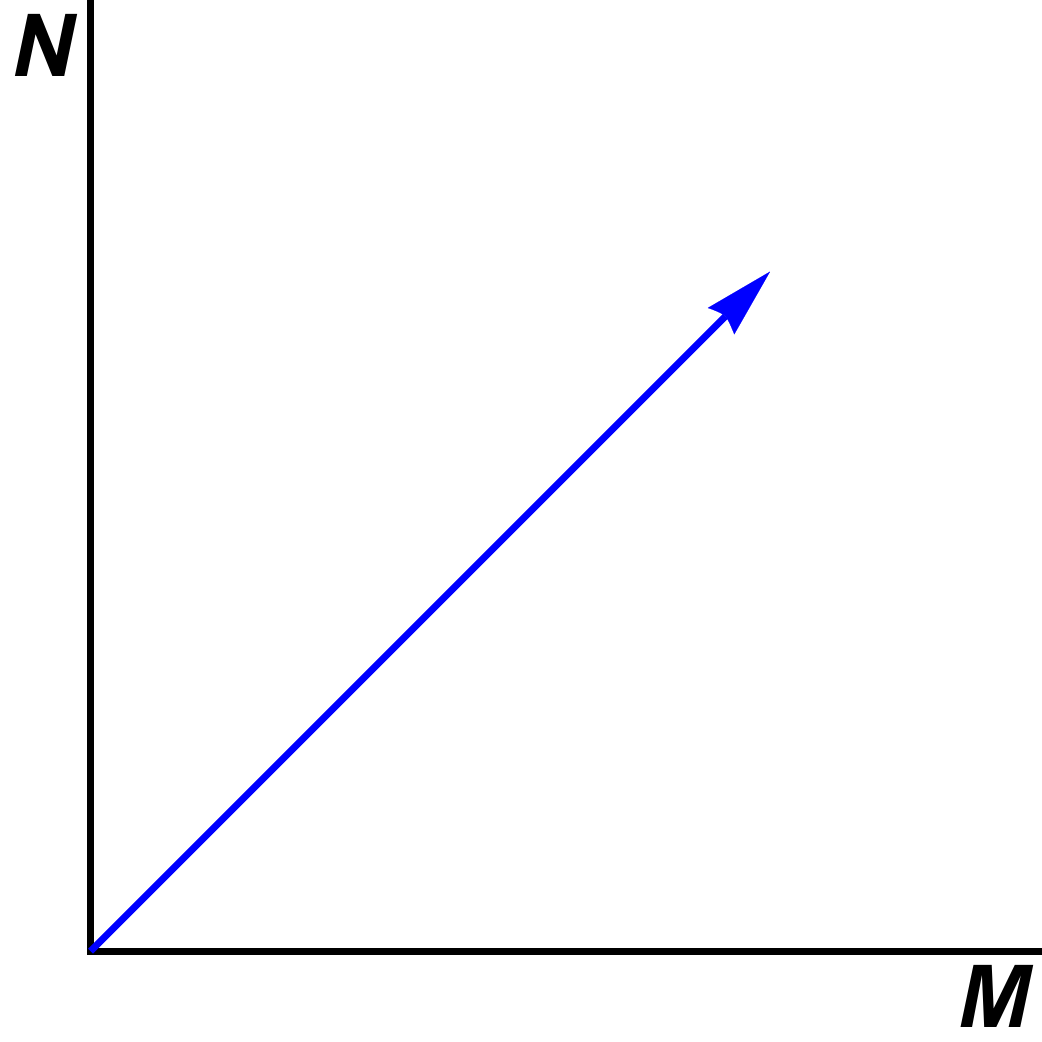}
}
\hspace{.1in}
\subfloat[$k=2$]{
\includegraphics[width=.3\textwidth]{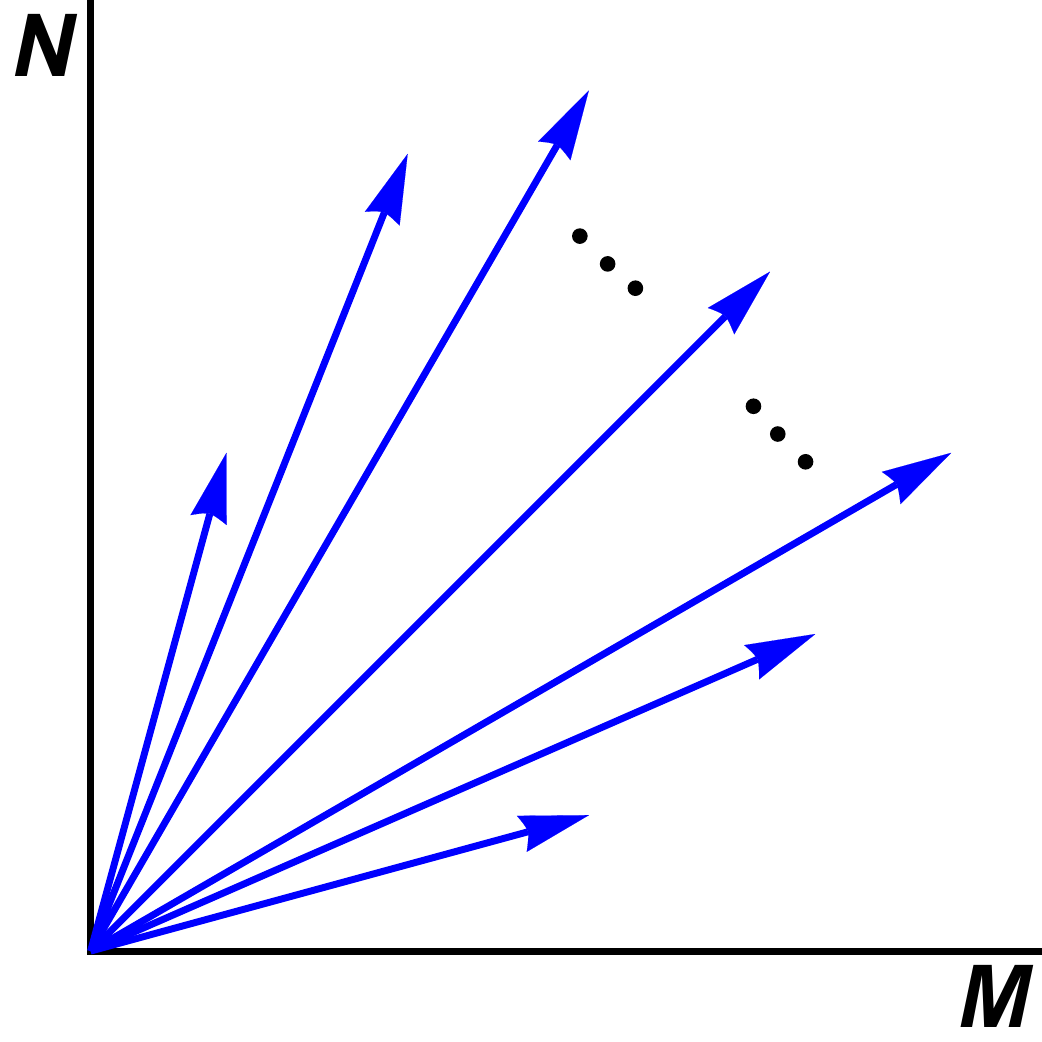}
}
\hspace{.1in}
\subfloat[$k>2$]{
\includegraphics[width=.3\textwidth]{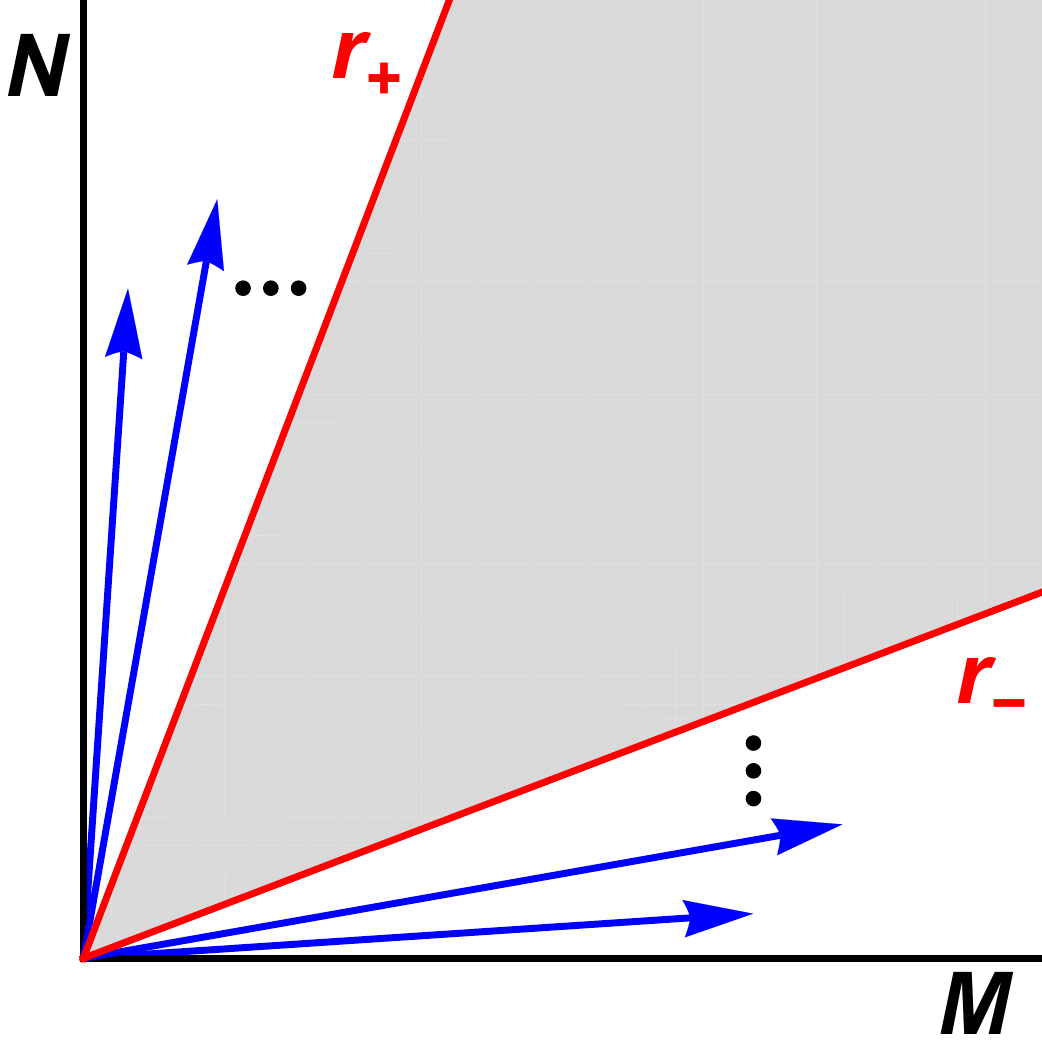}
}
\end{center}
\caption{ Occupied dimension vectors (i.e. pairs $(M,N)$) as a function of $k$.  In (a), the case $k=1:$ there is a single non-trivial dimension vector $(M,N)=(1,1)$.  In (b), the case $k=2:$ there are infinitely many occupied dimension vectors which accumulate at $r=1$.  In (c), the case $k>2:$  there are infinitely many occupied dimension vectors which accumulate along the irrational slopes $r=r_{\pm}.$  Inside the cone bounded by $r_{\pm}$ (shown in gray) the occupied dimension vectors are dense and the degeneracies grow exponentially.  }\label{fig:rays}
\end{figure}

Returning to the case of general ranks $M$ and $N$, we now summarize the qualitative possibilities for the large rank behavior of the degeneracies $\Omega(M,N,k)$ as a function of $k$. These behaviors are illustrated in Figure \ref{fig:rays}.\footnote{For all $k$ the degeneracies $\Omega(1,0,k)$ and $\Omega(0,1,k)$ are one and we do not discuss them further.}
\begin{itemize}
\item When $k=1,$ there is a single non-trivial degeneracy at $M=N=1$.  Thus, in this case there is no growth in the degeneracies for large ranks.  Physically, this model describes the BPS particles in the Argyles-Douglas conformal field theory \cite{Argyres:1995jj, Gaiotto:2009hg, Alim:2011ae}.
\item When $k=2,$ there are infinitely many non-trivial degeneracies, with allowed values $M=N\pm1$ and $M=N=1$.  In the former case the degeneracy is one, in the latter it is two.  Thus, again in this case there is no growth in the degeneracies for large ranks.  Physically, this model describes the BPS particles in the pure $su(2)$ Seiberg-Witten theory \cite{Seiberg:1994rs, Fiol:2000pd}.
\item When $k>2,$ there are infinitely many non-zero degeneracies.  Physically, this model occurs, for instance, as a subsector of $su(n)$ super Yang-Mills with $n>2$ \cite{Galakhov:2013oja}.  In general, there is no known closed form expression for the degeneracies, however previously known exact results from the case $N=M$ and $N=M+1$ indicate that the degeneracies grow exponentially for large ranks  \cite{Weist:2009, Galakhov:2013oja, Kim:2015oxa}.  

In this case, it is instructive to regard the degeneracies as a function of the limiting ratio $ N/M\rightarrow r$.  In terms of $r,$ the dimension of moduli space \eqref{dimform} reads 
\begin{equation}
\mathrm{dim}\left(\mathcal{M}^k_{M,N} \right)=kMN-M^{2}-N^{2}+1=M^{2}\big(kr-r^{2}-1\big)+\mathcal{O}(1/M)~. 
\end{equation}
The degeneracies can only be non-trivial if the above is non-negative.  For large $M,$ and fixed $r,$ this bounds the ratio $r$ between the two values $r_{\pm}$ given below 
\begin{equation}
r_{\pm} \equiv \frac{k\pm \sqrt{k^{2}-4}}{2}~.
\end{equation}
Inside the cone $r_{-}\leq r \leq r_{+},$ the occupied ratios are dense.  

Finally, we note the following inequalities which hold for $k>2$.
\begin{equation}
0 <r_{-}<1<k-1<r_{+}<k~.
\end{equation}
Thus, the interval $[r_{-}, r_{+}]$ contains $k-1$ integral values of $r$.  In \S \ref{sec:explicit}, we determine that the degeneracies also grow exponentially at these integral values of $r$.  

\end{itemize}

\subsection{Conjectured Asymptotics of $\Omega(M,N,k)$}
\label{sec:conjecture}

We now state a conjecture concerning the growth of the degeneracies $\Omega(M,N,k)$ for large ranks.  This conjecture was first articulated in \cite{weistthesis, Weist:2009}, and subsequently refined by \cite{Galakhov:2013oja}.

\textbf{Conjecture:}  For fixed $r, m, n,$ and $k>2,$ the degeneracies grow as follows
\begin{equation}
\frac{1}{M}\log\Bigg(\Omega(M+m,Mr+n,k)\Bigg)\underset{M\gg1}{\longrightarrow} S(r,k)+E(r,k,m,n)\frac{\log(M)}{M}+\cdots ~, \label{slopedef}
\end{equation}
where the terms $\cdots$ tend to zero faster than $\log(M)/M$ as $M$ tends to infinity.   

Let us expand upon several aspects of this conjecture.
\begin{itemize}
\item The leading asymptotics is controlled by the slope function $S(r,k)$ which is independent of the offset $(m,n)$.  Evidence for this independence can be given using explicit calculations from wall-crossing formulas and is presented in \S \ref{sec:wall}.
\item By contrast, the first correction to the leading growth, controlled by the function $E(r,k,m,n),$ depends on the offset $(m,n).$  This claim follows from known exact results for the degeneracies $\Omega(M,M+1,k)$ \cite{Weist:2009} and $\Omega(M,M,k)$ \cite{Reineke, Galakhov:2013oja}.  In these cases one finds
\begin{equation}
E(1,k,0,1)=-\frac{5}{2}~,\hspace{.5in}E(1,k,0,0)=-2~. \label{eexamp}
\end{equation}
\item The slope function $S(r,k)$ is assumed to be continuous on the interval $ r_{-}\leq r \leq r_{+}.$  Since the moduli spaces become empty at $r_{\pm}$ we have
\begin{equation}
S(r_{-},k)=S(r_{+},k)=0~.
\end{equation} 
For $r$ outside the interval $[r_{-},r_{+}],$ the slope function is not defined.
\item The leading growth implied by the conjecture is \emph{slower} than for generic quiver models.  In a generic quiver with node ranks $Q_{i}$ one expects that under scaling $Q_{i}\rightarrow \Lambda Q_{i},$ with $\Lambda\gg1$  the index $\Omega$ scales as $\log(\Omega)\propto \Lambda^{2}.$  Indeed, this is expected in quiver models that describe BPS black holes \cite{Denef:2007vg}.  By contrast, the Kronecker model, which occurs in quantum field theory, has  $\log(\Omega)\propto \Lambda.$

\end{itemize}
The slope function $S(r,k)$ is the primary quantity of interest in this work.  Assuming the validity of the conjecture, we constrain its functional form in \S \ref{sec:const}. In \S \ref{sec:explicit} we present calculations of the slope at integral values of $r$. 

\subsection{Constraints on the Slope Function}
\label{sec:const}

There are a number of a priori restrictions that may be put on the slope function $S(r,k)$ using dualities and known exact results.  We survey these constraints in this section.

\paragraph{Value at $r=1$}\mbox{}\\

The first piece of information about the slope, is that it is known exactly at the special value $r=1$.  Indeed, from \cite{Weist:2009}, we have the closed form expression
\begin{equation}
\Omega(M,M+1,k)= {k\over (M+1) \left[(k-1)M +k\right] }{ (k-1)^2 M +k(k-1) \choose M}~.
\end{equation}
This exact result is unusual.  For the majority of indices $\Omega(M,N,k)$ there is no simple known closed form expression.  Given this expression for finite $M,$ we may easily obtain its asymptotics for large $M$ using the Stirling approximation.  We find
\begin{equation}
S(1,k)= (k-1)^{2}\log((k-1)^{2})-(k^{2}-2k)\log(k^{2}-2k)~. \label{value}
\end{equation}

\paragraph{Reflection Symmetry}\mbox{}\\

\begin{figure}[here!]
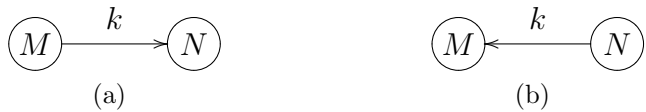

  \centering
\subfloat[]{
\xy  0;<1pt,0pt>:<0pt,-1pt>::
(-300,0) *+{M}*\cir<10pt>{} ="1",
(-240,0) *+{N}*\cir<10pt>{} ="2",
(-270, -10) *+{k} ="b",
\ar @{->} "1"; "2"
\endxy}
\hspace{1in}
\subfloat[]{
\xy  0;<1pt,0pt>:<0pt,-1pt>::
(-300,0) *+{M}*\cir<10pt>{} ="1",
(-240,0) *+{N}*\cir<10pt>{} ="2",
(-270, -10) *+{k} ="b",
\ar @{->} "2"; "1"
\endxy}
  \caption{The reflection symmetry.  In (a) the original model.  In (b) the quiver obtained after changing the definition of chiral and antichiral fields.  This operation replaces $\Phi_{i}$ with $\Phi_{i}^{\dagger}$ and hence reverses the arrows. }\label{fig:chiral}
\end{figure}

We may constrain the slope function $S(r,k)$ using symmetries of the quiver quantum mechanics.  One simple symmetry is that our choice of which fields we refer to as chiral and which fields refer to as antichiral is arbitrary.  Exchanging these notions changes the fields $\Phi_{i}$ to $\Phi_{i}^{\dagger},$ and hence reverses the direction of the arrows as shown in Figure \ref{fig:chiral}.  

It is clear that the net result of this operation is to exchange the roles of $M$ and $N$ in the definition of the index.  Thus, we have the symmetry
\begin{equation}
\Omega(M,N,k)=\Omega(N,M,k)~.
\end{equation}
We may translate this into a constraint on the slope function by using the definition \eqref{slopedef}.  We obtain
\begin{equation}
S(r,k)= r S(1/r,k)~.\label{refs}
\end{equation}

\paragraph{Mutation Symmetry}\mbox{}\\

\begin{figure}[here!]
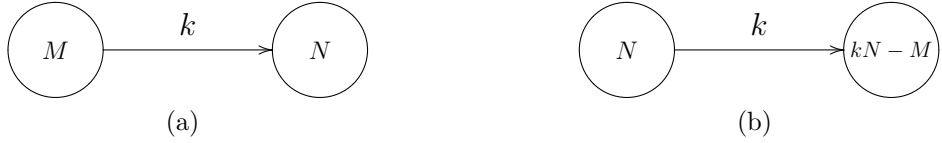

  \centering
\subfloat[]{
\xy  0;<1pt,0pt>:<0pt,-1pt>::
(-300,0) *+{\scalebox{.8}{$M$}}*\cir<18pt>{} ="1",
(-200,0) *+{\scalebox{.8}{$N$}}*\cir<18pt>{} ="2",
(-250, -10) *+{k} ="b",
\ar @{->} "1"; "2"
\endxy}
\hspace{1in}
\subfloat[]{
\xy  0;<1pt,0pt>:<0pt,-1pt>::
(-300,0) *+{\scalebox{.8}{$N$}}*\cir<18pt>{} ="1",
(-200,0) *+{\scalebox{.7}{$kN-M$}}*\cir<18pt>{} ="2",
(-250, -10) *+{k} ="b",
\ar @{->} "1"; "2"
\endxy}
  \caption{The mutation symmetry.  In (a) the original model.  In (b) the quiver obtained after a mutation. }\label{fig:mutation}
\end{figure}

A less trivial symmetry of the slope function follows from the application of quiver mutation (Seiberg dualities) \cite{MR0332887,MR0393065}.  Applying this operation enables us to change the ranks of the gauge groups in a $k$ dependent way as illustrated in Figure \ref{fig:mutation}.

The result of the mutation symmetry is thus to exchange $(M,N)\rightarrow (N,kN-M)$.  Correspondingly, we have symmetry
\begin{equation}
\Omega(M,N,k)=\Omega(N,kN-M,k)~.
\end{equation}
The resulting symmetry of the slope is 
\begin{equation}
S(r,k)=rS(k-1/r,k)~.\label{muts}
\end{equation}

\subsubsection{Solving the Constraints}
\label{sec:solve}

The totality of these constraints on the slope motivates us to introduce a function $G(r,k)$ and express the slope function as follows
\begin{equation}
S(r,k)=\Bigg(\sqrt{\frac{kr-r^{2}-1}{k-2}}\Bigg)\Bigg( (k-1)^{2}\log((k-1)^{2})-(k^{2}-2k)\log(k^{2}-2k)\Bigg)G(r,k)~. \label{gdef}
\end{equation}
To understand the significance of this formula, first note that the factor in the square root satisfies the algebraic identities 
\begin{equation}
\sqrt{kr-r^{2}-1} =r\sqrt{\frac{k}{r}-\frac{1}{r^{2}}-1}=r\sqrt{k\left(k-\frac{1}{r}\right)-\left(k-\frac{1}{r}\right)^{2}-1} ~.
\end{equation}
Therefore, the complete list of constraints on the function $S(r,k)$ translates into the following constraints on the quantity $G(r,k).$
\begin{itemize}
\item From the special value of the slope, \eqref{value}, we have
 \begin{equation}
 G(1,k)=1~.
 \end{equation}
\item From the reflection symmetry, \eqref{refs}, we have
\begin{equation}
G(r,k)=G(1/r,k)~.
\end{equation}
\item From the mutation symmetry, \eqref{muts}, we have
\begin{equation}
G(r,k)=G(k-1/r,k)~.
\end{equation}
\end{itemize}
Thus, assuming that the conjecture \eqref{slopedef} is true, it remains to find the function $G(r,k)$ which determines the value of the slope away from the special case $r=1$.  In \S \ref{sec:explicit} we provide direct calculations illustrating that the function $G(r,k)$ is not constant.  
In the remainder of this section, we continue to study its features by exploring the above constraints.

The functional identities obeyed by $G(r,k)$ may be viewed as fractional linear transformation acting on the variable $r.$  Specifically, given any $GL(2,\mathbb{Z})$ matrix, $X,$ define its action on $r$ in the standard way as
\begin{equation}
 X\cdot r= \frac{ar+b}{cr+d}~, \hspace{.5in}X=\left(\begin{array}{cc}a & b \\c&d \end{array}\right)~.
\end{equation}
The reflection and mutation symmetries are defined by the two $GL(2,\mathbb{Z})$ matrices
\begin{equation}
A=\left(\begin{array}{cc}0 & 1 \\ 1 & 0 \end{array}\right)~, \hspace{.5in}B=\left(\begin{array}{cc}k & -1 \\ 1 & 0 \end{array}\right)~. \label{ABdef}
\end{equation}
Our constraints on the function $G(r,k)$ may thus be rephrased by saying that $G(r,k)$ is a modular function for the subgroup of $GL(2,\mathbb{Z})$ generated by \eqref{ABdef}.

To understand the implications of the modular invariance of the function $G(r,k)$ it is useful to change coordinates from $r$ to a variable where the modular constraints are manifest.  An appropriate coordinate may be deduced by diagonalizing the mutation matrix $B$ above.  Upon defining $\theta$ as
\begin{equation}
\theta \equiv \frac{2\pi}{\log\left(r_{+}/r_{-}\right)}\log\left(\frac{r-r_{-}}{r_{+}-r}\right)~, \label{thetadef}
\end{equation}
we find that the transformations act simply as
\begin{equation}
(B\circ A)\cdot \theta =-\theta ~, \hspace{.5in}B \cdot \theta= \theta +2\pi~.
\end{equation}
Therefore, the constraints on the function $G(r,k)$ may be solved by expressing $G(r,k)$ in terms of the variable $\theta$ and demanding that it is even and periodic
\begin{equation}
G(\theta,k)=G(-\theta,k)=G(\theta+2\pi,k)~.
\end{equation}

Let us comment further on the coordinate transformation \eqref{thetadef}.  This transformation maps the segment $[r_{-},r_{+}]$ to the full real line $(-\infty,\infty).$ In particular the $r$ values $r_{\pm}$ map to the $\theta$ values $\pm \infty$. The fact (demonstrated in \S \ref{sec:explicit}) that $G(\theta,k)$ is not constant, implies that $G(\theta,k)$ undergoes infinitely many oscillations as $|\theta|$ increases.  Viewed in the original $r$ coordinate, these are oscillations with increasing frequency as $r$ approaches $r_{\pm}$.  

As a consequence of these considerations, we see that any non-constant $G(r,k)$ has the feature that its limit as $r\rightarrow r_{\pm}$ does not exist. Hence $G(r,k)$ is not continuous at the edges $r_{\pm}$ of the interval  $[r_{-},r_{+}]$ where the slope is defined.  This lack of continuity of $G(r,k)$ does not affect the claim that the full slope function $S(r,k)$ is continuous. Indeed, from \eqref{gdef} we see that the square root factor vanishes at $r_{\pm}$ so for continuity of the full slope it is sufficient that 
\begin{equation}
\lim_{r\rightarrow  r_{\pm}}G(r,k)\sqrt{kr-r^{2}-1}=0~. \label{conteq}
\end{equation}
In fact, we will see that $G(r,k)$ oscillates in a bounded range, so that the above is obeyed.

\section{Explicit Calculations of the Slope}
\label{sec:explicit}

In this section we provide new explicit calculations of the slope function $S(r,k).$  These calculations are possible due to new expressions for the degeneracies $\Omega(M,N,k)$ in the special case where $N=Mr+1$ for integral $r$.  To describe these results it is convenient to first introduce a generating function 
\begin{equation}
F(k,r,x) =(k-r) \sum_{\ell=1}^\infty  {(-1)^{\ell-1}\over \ell} {k\ell \choose r\ell }  x^\ell~, \label{fdef}
\end{equation}
and let $[x^j]\{ q(x) \}$ denote the coefficient of $x^{j}$ in a power series  $q(x)$. Then the result of \cite{Cordova:2015qka} is
\begin{equation}
\Omega(M,Mr+1,k)= {1\over (Mr+1)^2}[x^M] \left\{\exp\Big[\, (Mr+1) F(k,r,x)\, \Big] \right\}  ~. \label{result}
\end{equation}

In this section we use this expression to compute the slope $S(r,k)$ for the integral points $r=1,\cdots, k-1.$  In \S \ref{sec:saddle} we describe the saddle point technique for extracting the slope $S(r,k)$ from \eqref{result}.  In \S \ref{sec:limit}, we describe results for the slope function in limits where $k$ is also taken to be large.

\subsection{Saddle Point Approximation}\label{sec:saddle}

We begin by noting that \eqref{result} is equivalent to an expression for the degeneracy $\Omega(M,Mr+1,k)$  as a contour integral around $x=0$:\footnote{Generally, the function $\exp \left[ (Mr+1) F(k,r,x)\right]$ has a branch cut on the complex plane away from the origin. We  choose the radius $R$ of the contour integral  to be sufficiently small to avoid crossing the branch cut.}
\begin{align}\label{contour1}
\Omega(M,Mr+1,k)={1\over (Mr+1)^2}\oint_{x=0} {dx \over (2\pi i) x^{M+1} } \exp \Big[ \, (Mr+1) F(k,r,x) \, \Big]~.
\end{align}
Let us define the angular coordinate $\phi$ by $x=R e^{i\phi}$, where $R$ is the radius of the contour. In terms of $R$ and $\phi$, \eqref{contour1} can be expressed as
\begin{align}\label{contour2}
\Omega(M,Mr+1,k)={1\over (Mr+1)^2} {1\over R^M}\int_{0}^{2\pi} {d\phi\over 2\pi}
\exp \Big[  - i M\phi + (Mr+1) F(k,r,R e^{i\phi}) \Big]~.
\end{align}
When $M$ is very large, this integral is well approximated by the saddle point method. 

We now find the saddle point of \eqref{contour2} on the complex $\phi$ plane. Denote the saddle point by $\phi_s\in \mathbb{C}$ and define 
\begin{align}
x_s \equiv R \, e^{i\phi_s}~.
\end{align}
 The saddle point equation  is given by
\begin{align}\label{saddle}
{M\over Mr+1} = x_s \left.{d\over dx} F(k,r,x) \right\vert_{x=x_s}~.
\end{align}
Given the explicit power series expansion for $F(k,r,x)$, the saddle point equation can be solved to arbitrary numerical precision for any given $k$ and $r$. We make the following claim
\begin{align*}
\textbf{Claim:}~~&\text{The solution $x=x_s$ to \eqref{saddle} has a well-defined limit as $M\rightarrow \infty$} \\
& \text{for all $k>2$ and all integral $r$ with $1\leq r\leq k-1$.}
\end{align*}
This claim is justified by extensive numerical evidence.

Assuming this claim, we can rewrite the saddle point equation \eqref{saddle} as
\begin{align}\label{saddle2}
{1\over r}  = x_s \left.{d\over dx} F(k,r,x) \right\vert_{x=x_s} ~.
\end{align}
The index can be approximated by evaluating the integrand in \eqref{contour2} at $x_s$  in the large $M$ limit:
\begin{align}
\, \log \Omega(M,Mr+1,k) \simeq 
 M\Big[-  \log (x_s) + r F(k,r,x_s) \,\Big]+\mathcal{O}(\log (M)) \, ~,
\end{align}
We have therefore obtain the exponential growth of the index $\Omega(M,Mr+1,k)$ in the large $M$ limit.  Moreover, the slope function $S(r,k)$ is determined to be
\begin{align}\label{slopefunction}
S(r,k ) = -  \log (x_s) + r F(k,r,x_s)~,~~1\le r\le k-1,~~r\in \mathbb{N}~.
\end{align}
with $x_s$ defined as the solution to \eqref{saddle2}.

We can also give an exact expression for the function $G(r,k)$ defined in \eqref{gdef} for these values of $r$ simply by taking ratios,
\begin{align}\label{Gfunction}
\begin{split}
G(r,k) &= \sqrt{k-2\over kr -r^2-1}{ -  \log (x_s) + r F(k,r,x_s) \over (k-1)^2 \log \left[ (k-1)^2\right] -(k^2-2k)\log (k^2-2k) }~,\\
&~~~~~~~~~~~~~~~~~~~~~~~~~~~~~~~~~~~~~~~~~~~~~~~~~~~~~~~~~~~~~~~~~~1\le r\le k-1,~~r\in \mathbb{N}~.
\end{split}
\end{align}

Given the explicit form of the function $F(k,r,x)$ \eqref{fdef}, the saddle point equations \eqref{saddle2}-\eqref{Gfunction} may be solved to arbitrary numerical precision.   Using the symmetries of the slope function discussed in \S \ref{sec:solve} we may then extrapolate these results to larger and smaller non-integral values of $r.$  Interpolating between these data points (assuming continuity of $S(r,k)$) then provides a plausible picture of the slope for all $r$ in the interval $[r_{-},r_{+}].$  We present such plots in Figures \ref{fig:s} and \ref{fig:g} below.  Note that $G(r,k)$ oscillates and $S(r,k)$ goes to zero as $r\rightarrow r_{\pm}$ as anticipated in \eqref{conteq}.

\begin{figure}[h!]
\centering
\raisebox{6.5pt}{\includegraphics[width=.36\textwidth]{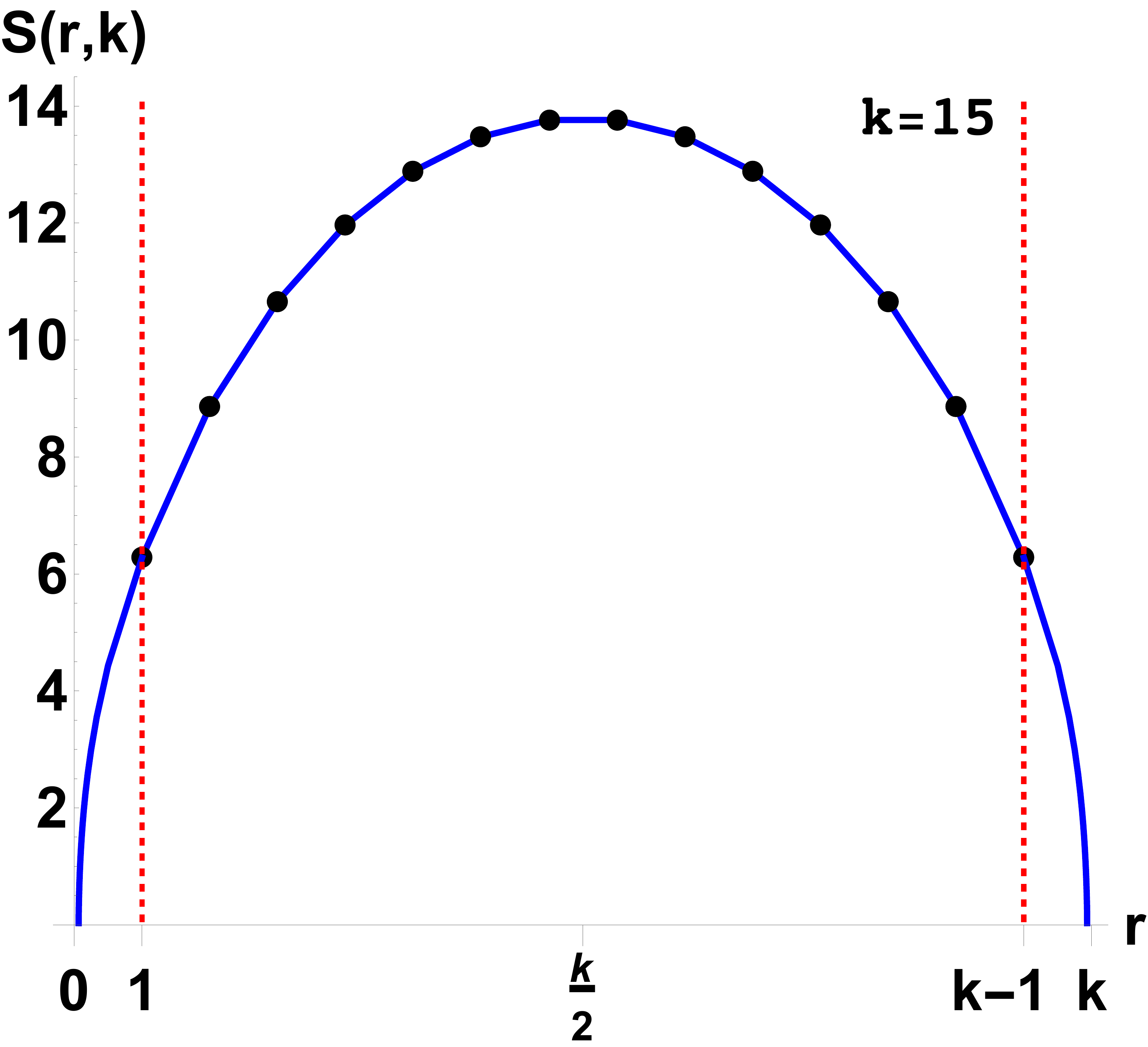}}
~~~
\includegraphics[width=.6\textwidth]{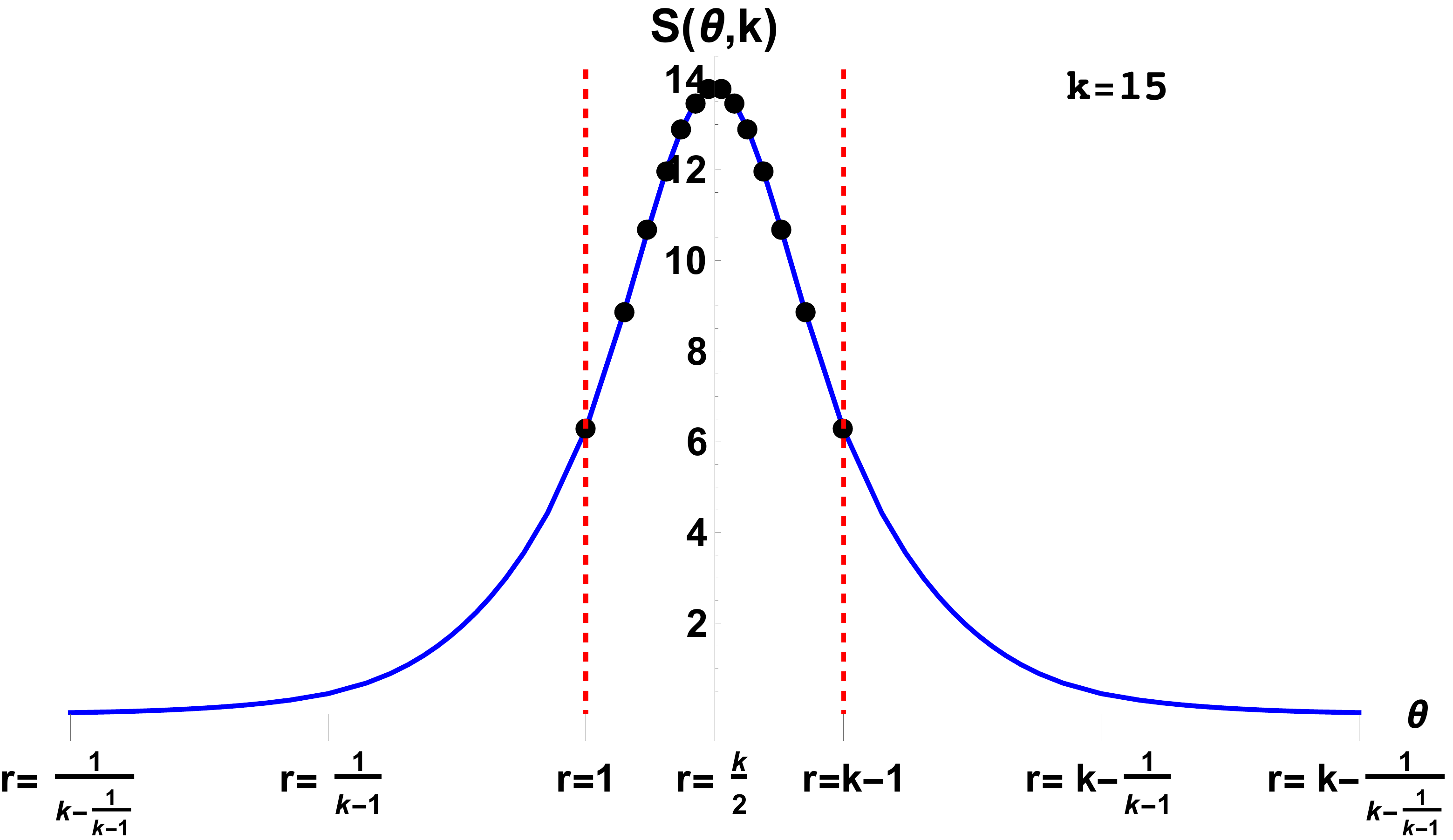}
 \caption{The slope function $S(r,k)$ in the case $k=15$.   On the left the independent variable is $r,$ on the right the independent variable is $\theta.$ The marked points denote the values of the slope computed using the saddle point method.  These points may be transferred to $r<1$ and $r>k-1$ (outside the red dashed lines) using the symmetries of the slope function. The blue curve is the resulting interpolating function. }\label{fig:s}
\end{figure}

\begin{figure}[h!]
\raisebox{5pt}{\includegraphics[width=.38\textwidth]{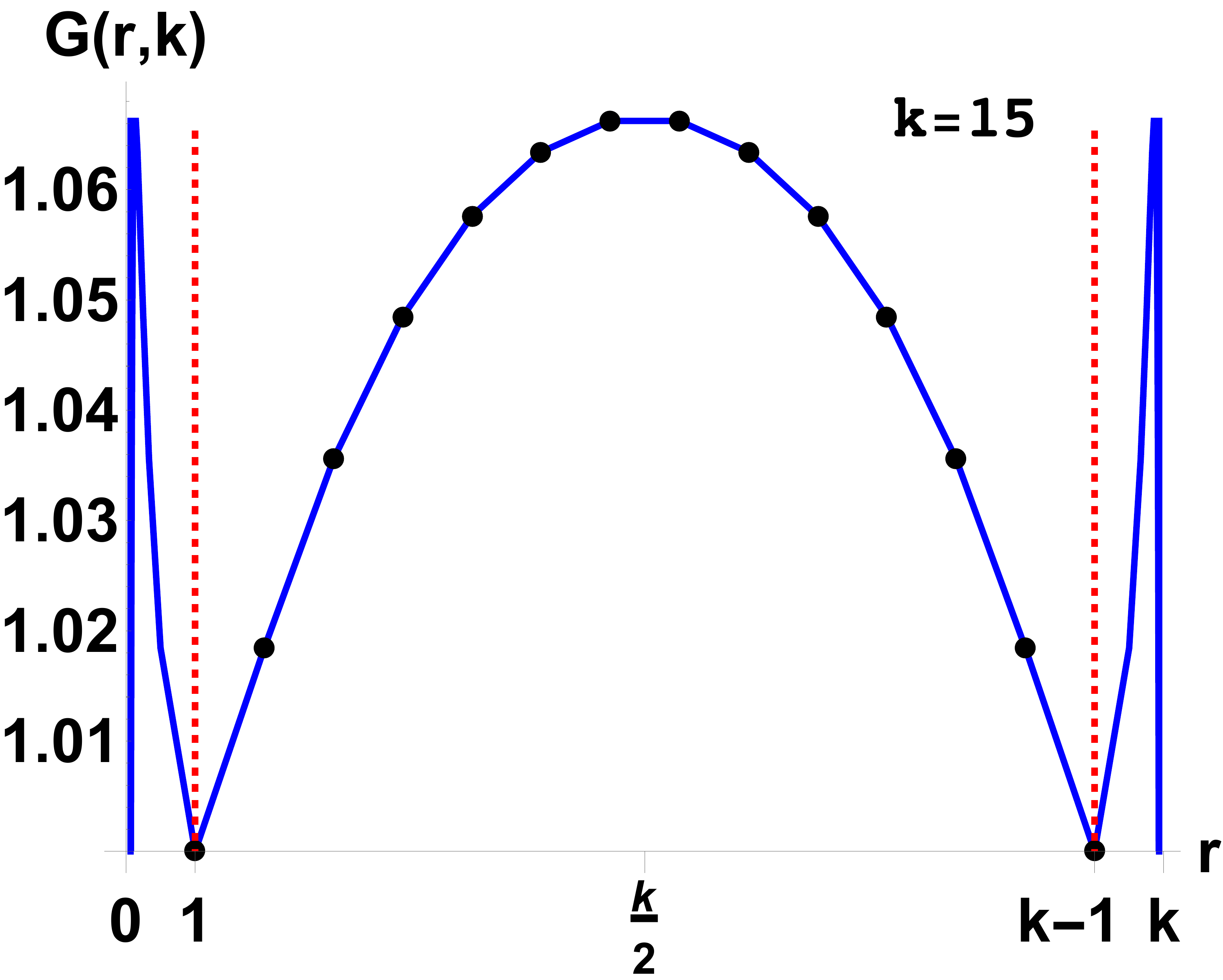}}
~~~
\includegraphics[width=.6\textwidth]{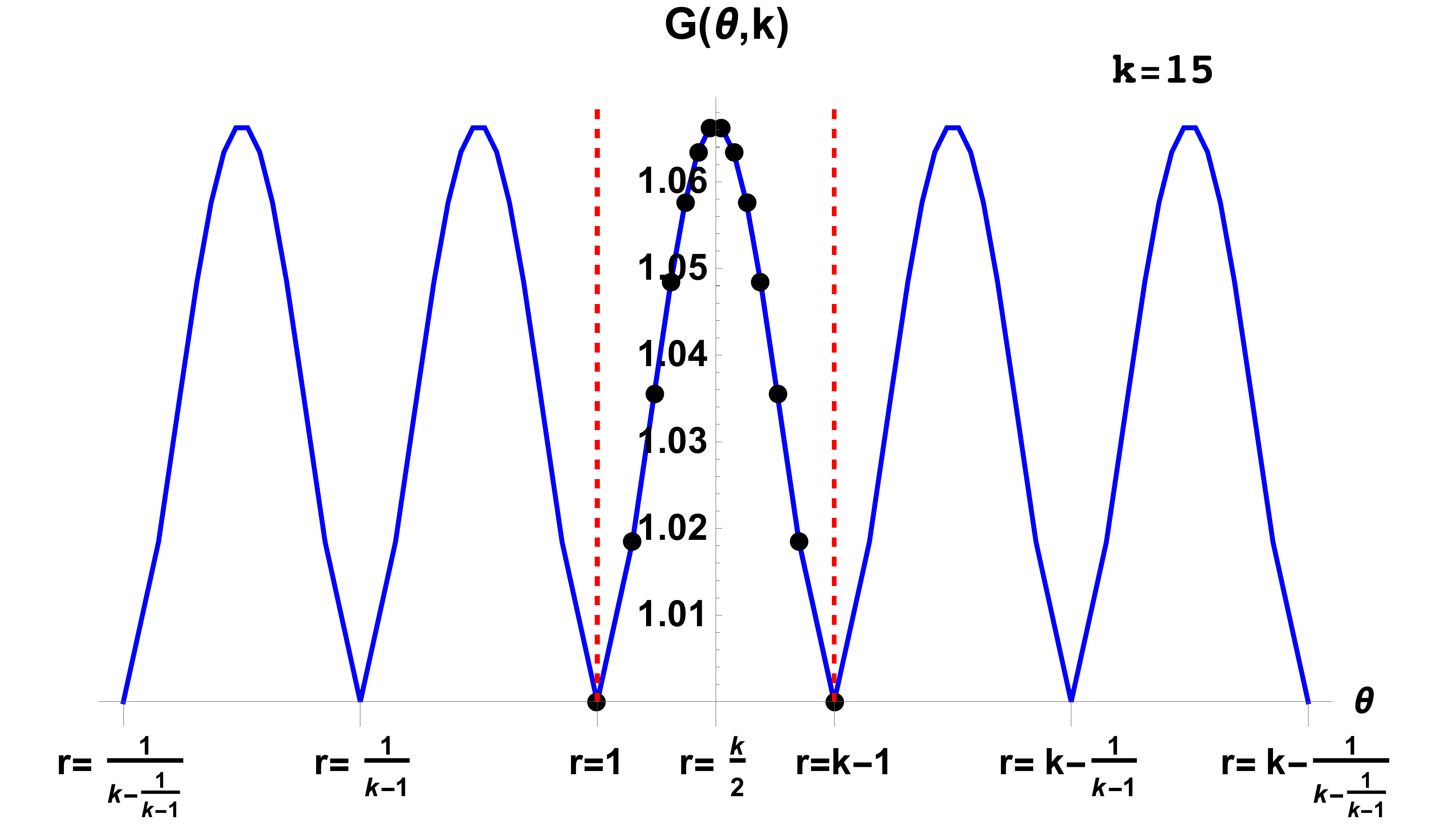}
 \caption{The slope function $G(r,k)$ in the case $k=15$.   On the left the independent variable is $r,$ on the right the independent variable is $\theta.$ The marked points denote the values of the slope computed using the saddle point method.  These points may be transferred to $r<1$ and $r>k-1$ (outside the red dashed lines) using the symmetries of the slope function. The blue curve is the resulting interpolating function.  Note that $G(r,k)$ undergoes infinitely many oscillations for $r_{-}<r<1$ and $k-1<r<r_{+}$.}\label{fig:g}
\end{figure}

\subsubsection{The Subleading Term}

The explicit expression \eqref{result} and the saddle point analysis also enables us to study the subleading $\log (M)$ term in $\log \Omega(M,Mr+1,k)$. This term receives two contributions: one from the $1/(Mr+1)^2$ term in \eqref{contour2}, and the other from the ``one-loop" correction from the integrating out the $\delta\phi^2$ term when expanding around the saddle point $\phi=\phi_s+\delta\phi$. Together they give
\begin{align}
\, \log \Omega(M,Mr+1,k) \simeq 
M\Big[-  \log (x_s) + r F(k,r,x_s) \,\Big]- {5\over2} \log( M)  + \mathcal{O}(1)\, ~.
\end{align}
This determines the function $E(r,k,m,n)$ appearing in \eqref{slopedef} for this particular value of the offset $(m,n)=(0,1)$:
\begin{align}
E(r,k,0,1) = - {5\over2}~,~~1\le r\le k-1,~~r\in \mathbb{N}~,
\end{align}
which generalizes the result \eqref{eexamp} of \cite{Galakhov:2013oja} to general integral $r$.
\subsubsection{Symmetry of the Slope Function}

Finally, we can also use saddle point analysis to check some of the symmetries of the slope function $S(r,k)$ that we argued for on general grounds in \S \ref{sec:const}.  Since our saddle point analysis is only valid for integral values of $r$, the only symmetry we can check is the composition of the mutation  and the  reflection symmetry:
\begin{align}\label{slopesym}
S(r,k) = S( k-r,k)~, \longleftrightarrow S(\theta,k)=S(-\theta,k)~.
\end{align}
To illustrate this result, we first note from \eqref{fdef} that
\begin{align}
F(k,k-r,x) = {r\over k-r} F(k,r,x)~.
\end{align} 
In other words, the combination $r\, F(k,r,x)$ is invariant under the symmetry \eqref{slopesym} $r\rightarrow k-r$. Since both the saddle point equation \eqref{saddle2} and \eqref{slopefunction} depend on $F(k,r,x)$ only through the combination $r\,F(k,r,x)$, it follows that the slope function $S(r,k)$ given in \eqref{slopefunction} indeed enjoys the symmetry \eqref{slopesym}. This reflection symmetry is manifest in Figures \ref{fig:s} and \ref{fig:g}.

\subsection{Limits of the Slope Function}\label{sec:limit}

In this subsection we further take limits on $k$ and $r$ to explore the behavior of $S(r,k)$ in different regimes of parameters. We emphasize that, in all such calculations, we  first take the large $M$ limit, and then take further limits on $k$ and $r$.

\subsubsection{Large $k$ with Fixed $r$}

We begin with the limit:
\begin{align}
k\rightarrow \infty,~~~r=\text{fixed}~.
\end{align}
Using the Stirling approximation, $n! \simeq n^n e^{-n} \sqrt{2\pi n}$, we can rewrite the saddle point equation \eqref{saddle2} as
\begin{align}\label{largeksaddle}
{1\over r} \simeq k\sqrt{k\over 2\pi r (k-r)}\sum_{\ell=1}^\infty {(-1)^{\ell -1}\over \sqrt{\ell}} \left[ { k^k \over r^r (k-r)^{k-r}  }x_s\right]^\ell~.
\end{align}
To solve the saddle point equation in this limit, we  truncate the righthand side to the first term $\ell=1$. The saddle point $x_s$ in the large $k$ limit is then given by
\begin{align}
x_s \simeq k^{-{3\over2}}\sqrt{2\pi (k-r)\over r} {  r^r (k-r)^{k-r}\over k^k  }~.
\end{align}
As a consistency check on our truncation to the $\ell=1$ term in the saddle point equation \eqref{largeksaddle}, we note that the $\ell$-th order term on the righthand side of \eqref{largeksaddle} evaluated at the saddle is
\begin{align}
 k\sqrt{k\over 2\pi r (k-r)} {(-1)^{\ell -1}\over \sqrt{\ell}} \left[k^{-{3\over2}} \sqrt{2\pi (k-r)\over r}\right]^\ell \sim k^{1-\ell}~.
\end{align}
Hence the terms with $\ell>1$ are suppressed and our truncation to $\ell=1$ is self-consistent in the large $k$ limit.

Given the explicit expression for the saddle point $x_s$ at large $k$, we can now solve for the slope function $S(r,k)$  \eqref{slopefunction} we obtain,
\begin{align}
\begin{split}
S(r,k)\xrightarrow{k\gg 1}
  (r+1)\log k + \left[ r- \log\left(\sqrt{2\pi} r^{r-{1\over2}}\right)+1\right] +\mathcal{O}\left({1\over k}\right)~.
 \end{split}
\end{align}
Note that in the large $k$ limit the dominant contribution comes from $-\log (x_s)$ in \eqref{slopefunction}. From this we also obtain the large $k$ limit of the function $G(r,k)$ \eqref{Gfunction},
\begin{align}
\lim_{k\rightarrow \infty}G(r,k) =  {r+1\over 2\sqrt{r} }~.
\end{align}
These results may be phrased simply in terms of the original degeneracy $\Omega(M,N,k)$ as
\begin{equation}
\lim_{k\rightarrow \infty} \lim_{\substack{M, N\rightarrow \infty \\ N/M=r \ \text{fixed}}} \Omega(M,N,k) \approx k^{M+N}~.
\end{equation}

\subsubsection{Large $k$ and $r$ with  Fixed $r/k$}

As another accessible limit, consider the case where
\begin{align}
k,r\rightarrow \infty,~~~q:= {r\over k} =\text{fixed}~. 
\end{align}
The constraint $1 \le r \le k-1$ becomes in this limit
 \begin{align}
0 \le q\le1~.
 \end{align} 
Again using the Stirling approximation, the saddle point equation \eqref{saddle} can be written as
\begin{align}\label{kr}
{1\over kq} \simeq  \sqrt{k}  \sqrt{1-q\over 2\pi q} \sum_{\ell=1} {(-1)^{\ell-1}\over\sqrt{\ell}}
\left[
{k^k \over (qk)^{qk} \left[ (1-q)k\right]^{(1-q)k} }x
\right]^\ell.
\end{align}
Upon truncating \eqref{kr} to the first term $\ell=1$, we obtain the saddle point
\begin{align}
x_s \simeq k^{-{3\over 2}} \sqrt{ 2\pi \over q(1-q)} { (qk)^{qk} \left[ (1-q)k\right]^{(1-q)k}\over k^k}~.
\end{align}
As a consistency check on our truncation to the $\ell=1$ term, we note that the $\ell$-th term on the righthand side of \eqref{kr} scales like $k^{ {1\over 2} - {3\ell\over 2}}$, which is negligible compared with the lefthand side when $\ell>1$.

Given the explicit expression for the saddle point $x_s$ at large $k$ and $r$ limit, we can then solve for the slope function $S(r,k)$ 
\begin{align}
\begin{split}
S(r,k) \underset{q\equiv r/k=\text{fixed}}{\xrightarrow{k,r\rightarrow \infty}}
-\Big[\, q\log (q)+  (1-q) \log(1-q) \,\Big] \,k +{3\over 2}\log k + \mathcal{O}(1)~. 
 \end{split}
\end{align}
In contrast to the large $k$ limit with $r$ fixed, the slope now scales linearly with $k$. 

\begin{figure}[h!]
\centering
\includegraphics[width=0.6\textwidth]{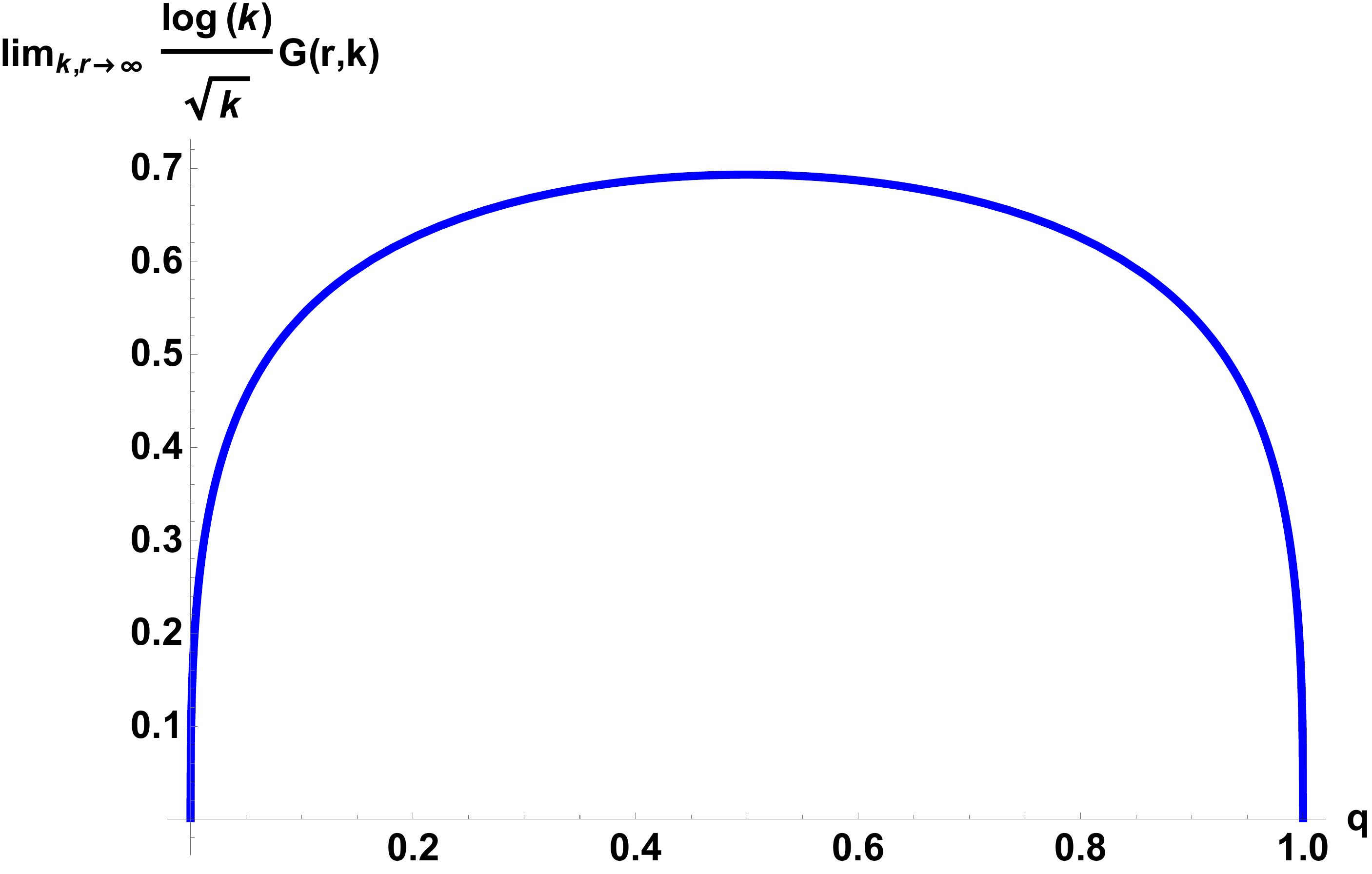}
\caption{The function $G(r,k)$ as a function of $q=r/k$ in the limit $k,r\rightarrow\infty$ with $q$ fixed. }\label{fig:Ginfinity}
\end{figure}

Meanwhile, the function $G(r,k)$ given by \eqref{Gfunction} behaves as
\begin{align}
\lim_{\substack{k,r\rightarrow \infty \\ q\equiv r/k=\text{fixed}}} G(r,k)
= - {  q\log (q)+  (1-q) \log(1-q) \over 2\sqrt{q(1-q) } }\, {\sqrt{k}\over \log  k} +\cdots~.
\end{align}
Thus, in this limit, $G(r,k)$  as a function of the ratio $q$ is symmetric under $q\rightarrow 1-q$ and has a maximum at $q=1/2$.  Note also that in this limit $G(r,k)$ grows in absolute value as $\sqrt{k}/\log(k).$  A plot of $G(r,k)$ in this regime of parameters is shown in Figure \ref{fig:Ginfinity}.

\section{Slopes from Wall-Crossing Data}
\label{sec:wall}

In this section, we describe the information that can be learned about the slope function $S(r,k)$ using data about the degeneracies obtained from the wall-crossing formula.  Our main goal is to provide evidence for an aspect of the conjecture stated in \S \ref{sec:conjecture}.  Namely, we wish to show that the slope function $S(r,k)$ defined as
\begin{equation}
\lim_{M\rightarrow \infty}\frac{1}{M}\log\Big(\Omega(M+m,Mr+n,k)\Big )= S(r,k)~,
\end{equation}
is indeed independent of the offset $(m,n)$.  Similar analysis has been preformed in \cite{Galakhov:2013oja}.

For general $(m,n),$ there is no known closed form expression for the indices which feature in the above.  Thus, it is presently impossible to conclusively prove or disprove the claim that $S(r,k)$ is independent of the offset $(m,n)$.  Instead, we can obtain evidence for this idea through explicit calculations of the degeneracies using wall-crossing.

The wall-crossing formula of \cite{Kontsevich:2008fj} enables us to find the change in $\Omega(M,N,k)$ as the Fayet-Iliopoulos parameters $\zeta$ are varied.  In the Kronecker model, the wall-crossing formula is straightforward to use.  If we change the sign of the FI parameter $\zeta$ of \eqref{dterm}, then all moduli spaces are empty.  Thus, in this simple chamber, the only values of $(M,N)$ with non-vanishing degeneracies are $(1,0)$ or $(0,1),$ corresponding to a single particle of type one, or a single particle of type two.  We therefore use this simple chamber ($\zeta<0$) as a seed, and use wall-crossing to determine the indices in the chamber of interest ($\zeta>0$) where the exponential growth in degeneracies occurs.

The wall-crossing calculation makes of functions $K_{M,N}$ defined as power series in formal variables $\Big[x,y\Big]$ as
\begin{equation}
K_{M,N}\Big[  x,y \Big]= \Big[\, x(1-(-1)^{kMN}x^{M}y^{N})^{kN}, y(1-(-1)^{kMN}x^{M}y^{N})^{-kM}\,\Big]~.
\end{equation}
Additionally, we define a sign function $\sigma$ that detects the parity of the dimension of $\mathcal{M}^{k}_{M,N}$
\begin{equation}
\sigma(M,N,k)=\begin{cases}+1~, &  kMN-M^{2}-N^{2}+1 \equiv 0 \  (\mathrm{mod} \ 2)~,\\
-1~, & kMN-M^{2}-N^{2}+1\equiv 1 \  (\mathrm{mod} \ 2)~.\end{cases} \label{signfunc}
\end{equation}

The content of the wall-crossing formula is that a certain function of $[x,y]$ built from compositions of the $K_{M,N}$ does not depend on the chamber.  In the Kronecker model this reads
\begin{equation}
\prod_{M,N \geq 0}^{\rightarrow}K_{M,N}^{\sigma(M,N,k)\Omega(M,N,k)} = K_{0,1} \circ K_{1,0}~. \label{wallcrossing}
\end{equation}
In the above, the product of operators $K_{M,N}\Big[  x,y \Big]$ is defined to be composition of functions, and the order of composition is that of decreasing $M/N.$\footnote{If $M_{1}/N_{1}=M_{1}/N_{2}$ then $K_{M_{1},N_{1}}\circ K_{M_{2},N_{2}}=K_{M_{2},N_{2}}\circ K_{M_{1},N_{1}}.$  The need for this sign $\sigma$ due to the fact that we have defined $\Omega$ to coincide with the Euler characteristic.}

To use \eqref{wallcrossing}, observe that $K_{M,N}$ differs from the identity first at order $x^{M}y^{N}.$  Therefore, fixing an integer $Q,$ we may solve \eqref{wallcrossing} to order $Q$ by truncating the infinite composition to a finite composition where only those $K_{M,N}$ are retained with $M+N\leq Q$.  Next we evaluate the composition as a polynomial by only retaining terms differing from the identity up to total order $Q$.  Matching to the right-hand side, we can then solve for all $\Omega(M,N,k)$ with $M+N\leq Q$.

This procedure is time consuming to carry out for large $Q$, and does not directly enable us to analytically determine a closed form expression for the slope function.  However, it does enable us to provide evidence for the claim that the slope is independent of the offset.  

\begin{figure}[h!]
\centering
\subfloat[]{
\includegraphics[width=.5\textwidth]{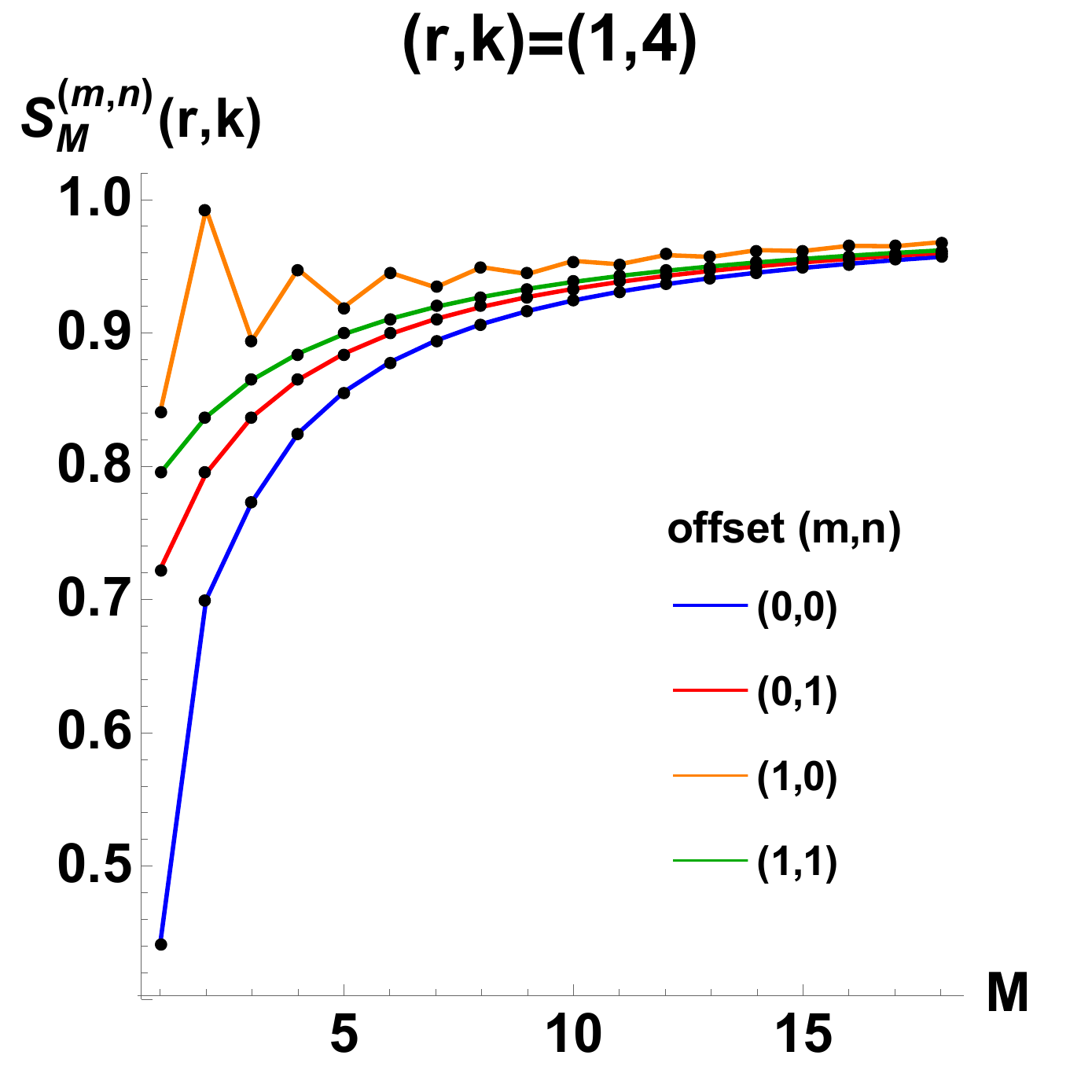}
}~~~~~
\subfloat[]{
\includegraphics[width=.5\textwidth]{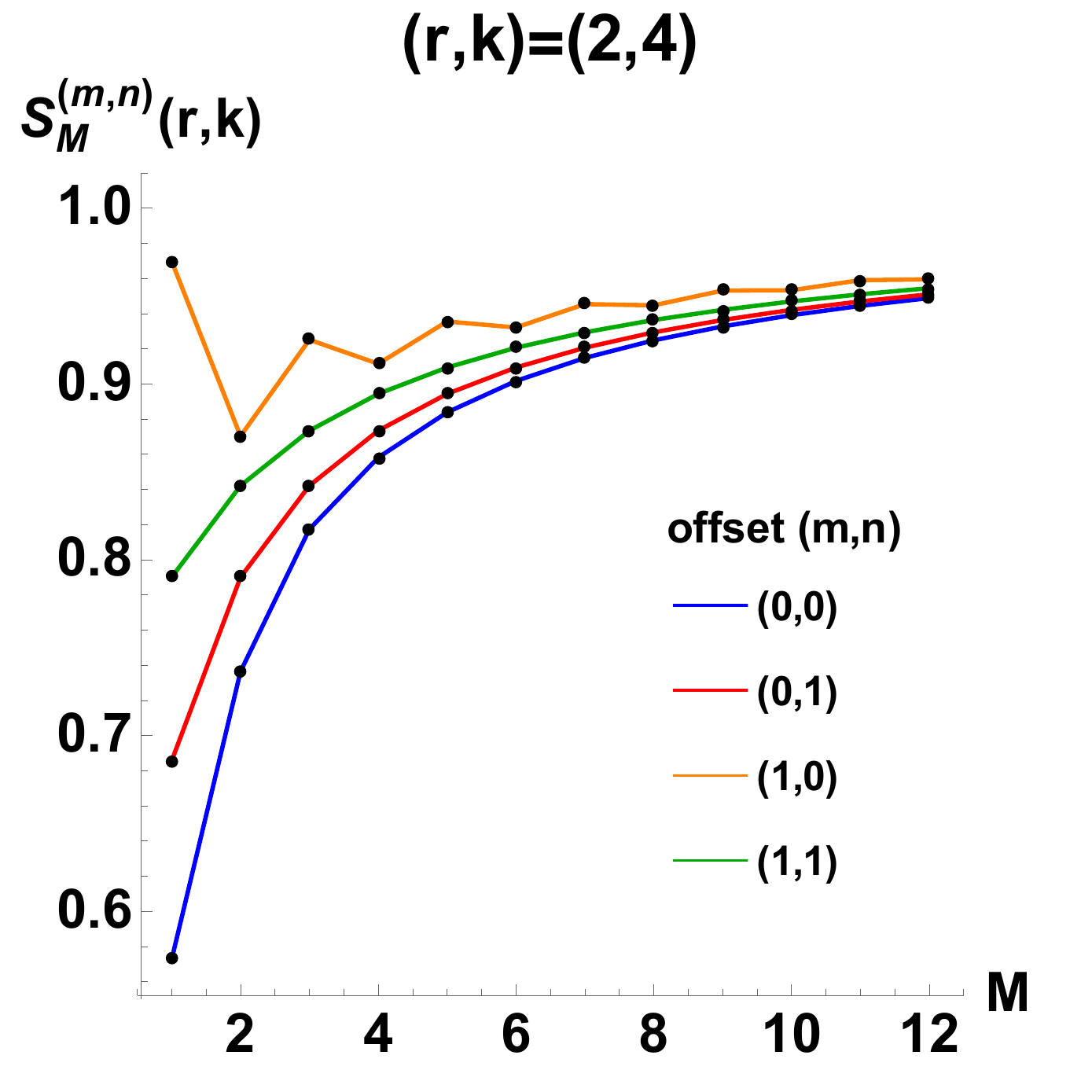}
}
\caption{ Examples of the normalized slope sequence $S_{M}^{(m,n)}(r,k)$ of \eqref{sequencedef} for various offsets at $k=4.$  (a) The case $r=1.$  (b) The case $r=2$. In both examples the convergence of the various curves suggests that asymptotic slope is independent of the offset.
 }
 \label{fig:wall}
\end{figure}

To do so, first define for each $(r,k),$ and each offset $(m,n),$ the following normalized sequence $S^{(m,n)}_{M}(r,k)$ 
\begin{equation}
S^{(m,n)}_{M}(r,k)\equiv \frac{\log \big[\Omega(M+1+m,(M+1)r+n,k)\big]-\log \big[\Omega(M+m,Mr+n,k)\big]}{S(r,k)}~. \label{sequencedef}
\end{equation}
For large $M,$ these sequences approximate a normalized version of the slope function.  Independence of the offset $(m,n)$ implies that the limit is unity
\begin{equation}
\lim_{M\rightarrow \infty}S^{(m,n)}_{M}(r,k)=1~. \label{limitconj}
\end{equation} 
We have studied these sequences using wall-crossing data (recorded in Appendix \ref{sec:walldata}).  Data collected thus far supports the result \eqref{limitconj}.  We illustrate examples in Figure \ref{fig:wall}.

\section{Algebraic Asymptotics}
\label{sec:alg}

In this section we explore the number theoretic properties of the slope function $S(r,k).$  Curiously, we observe that the exponential of the slope is an \emph{algebraic} number (i.e. solves a polynomial equation with integral coefficients) in all examples we have studied.  This leads us to conjecture the following:

\textbf{Conjecture:} For any rational $r$ with $r_{-}\leq r \leq r_{+}$, and any $k>2,$ the quantity $~~~~~~~~~~~~~~~~~~~~~~~~~$$\exp(S(r,k))$ is algebraic.

Before describing our method for verifying this conjecture at special values of $r$ and $k$, let us first describe what may be its physical content.  It has been observed in \cite{Kontsevich:2008fj, Gross, Reineke, Galakhov:2013oja, Galakhov:2014xba, Tom} that certain generating functions of threshold bound states obey algebraic equations.  

For an explicit example, consider the degeneracies  $\Omega(M,M,k).$  These ranks are not coprime and hence the quiver quantum mechanics is not gapped.  The ground states, counted by $\Omega(M,Mr,k)$ are thus at threshold.  We may assemble these degeneracies into a formal multiplicative generating function as
\begin{equation}
P_{k}(z)\equiv \prod_{\ell=1}^{\infty}\Big(1-(-1)^{\ell k}z^{\ell}\Big)^{\frac{\ell}{k} \sigma(\ell,\ell,k)\Omega(\ell,\ell,k)}~,
\end{equation}
where $\sigma$ is the sign function introduced in \eqref{signfunc}.  Then, remarkably, one finds that this generating function obeys the algebraic equation
\begin{equation}
P_{k}(z)=1+zP_{k}(z)^{(k-1)^{2}}~.
\end{equation} 
Algebraic equations, such as the above, suggest a combinatorial interpretation of threshold bound states.  Moreover, if such algebraic equations are a feature at general ratio $r$ (not just $r=1$) then they also provide evidence that $\exp(S(r,k))$ is indeed algebraic for general rational ratio.

In practice since we do not have access to such equations, our method for demonstrating that $\exp(S(r,k))$ is algebraic is less direct.  We carry out this analysis at integer $r$ where the saddle point approximation method of \S \ref{sec:explicit} can be applied.  Using this method we may evaluate the slope $S(r,k)$ to extremely high precision, say $p$ decimal digits.  With the aid of computer software,\footnote{Specifically, we use the ``RootApproximant" function in Mathematica.} we then ``guess" simple algebraic equations obeyed by the slope $S(r,k)$ to the given precision $p$.   We then test the validity of the resulting equations by evaluating their roots to precision $q > p$ and comparing against the numerical saddle value of the slope at the same higher precision $q$.  Agreement for large $q$ strongly suggests that we have hit upon the correct algebraic equation. 

We have carried out this algorithm for $r$ and $k$ sufficiently small.   In practice in these examples the precision $p$ used to determine the equation is of the order of $3000$ decimal digits, and the precision $q$ used to test the equation is of the order of $10000$ decimal digits, thus giving overwhelming evidence that the equations to follow are correct.  Remarkably, even for small values of these parameters, the resulting algebraic equations have large unfamiliar coefficients. We present examples of these polynomials below in the special case $r=2$ and increasing $k$.  In each case, $\exp(S(r,k))$ is the unique positive root of the given polynomial.  The complexity of these results demands explanation.   

\begin{align*}
&\bullet r=2,~~k=3:~~~~~~~~~~~~~~~~~~~~~~~~~~~~~~~~~~~~~~~~~~~~~~~~~~~~~~~~~~~~~~~~~~~~~~~~~~~~~~~~~~~~~~~~~~~~~~~~~~~~~\\
&256-27x~.\\
&\bullet r=2,~~k=4:~~~~~~~~~~~~~~~~~~~~~~~~~~~~~~~~~~~~~~~~~~~~~~~~~~~~~~~~~~~~~~~~~~~~~~~~~~~~~~~~~~~~~~~~~~~~~~~~~~~~~\\
& 600362847   + 440675453820928\, x 
- 8916100448256 \,x^2 ~. \\
&\bullet r=2,~~k=5:\\
&-591413771772821360012500490693032929265968209672451145145917265965744128+1544\\
&52605112448522226515494740065379723981012919983800320000000000000000000000000 \,x-\\
&316522677763135004318093459039662462828346178866922855377197265625000000000000000\\
&000000000 \,x^2+2350988701644575015937473074444491355637331113544175043017503412556\\
&834518909454345703125 \,x^3~.\\
&\bullet r=2,~~k=6:\\
&-2050773823560610053645205609172376035486179836520607547294916966189367296000000\\
&00000000000000+20352745636594082793019947349596049624338559361382715379523490900\\
&5427691102560816224960099284771703619584 \,x-64247493083782190701390594106712115785\\
&844636100938458366932530627926893892348597995911202127555619214492164722524160 \,x^2+\\
&2282730363469670449799005123371655224008190247224909338299547930732677173150041355\\
&90642802687246850771579138342847 \,x^3~.
\end{align*}
\begin{align*}
&\bullet r=2,~~k=7:\\
&-63221044749875358413745022419037852999211460218912354859182278723793194763608955\\
&69083158515272883214826719259981544144401779857393246859297454316875384045119131604\\
&09730106129249387945348177411325214212521132649260658490824379019540346646977013733\\
&70364308357238769531250000000000000000000000000000000000000000000000000000000000000\\
&00000000000000000000000000000000000000000000000000000000000+7689317001436808688465\\
&94496835357146564963658043009477662494651518387460225724968177988381514455526109803\\
&65480372933812665981956065023979490740085467402846935868178202385437133043353873557\\
&88224259870421298440415508596182706423035926152139981309326189119185155267175834498\\
&11677098750806180760264396667480468750000000000000000000000000000000000000000000000\\
&00000000000000000000000000000000000000000000000000\,x-11512622292097404993525570606\\
&31695997866311286244092918011731114409309096075272174764599912523483510094042695705\\
&33344879143879922000131490824383736952944636806713017012973184881908620479323776481\\
&32622103415724647423659250178192846527701864758976555597044399924916019009630666772\\
&76609369404201484050924563447725224030193658109055832028388977050781250000000000000\\
&000000000000000000000000000000000000000000000000000000000000\,x^2-672232996605797709\\
&29375864747083052183628238733887917868657258005534772442463791434241140103111792486\\
&89686734542268517271729316048108628676198322579421957110454083577542901318435903470\\
&04907611409361306027920753596289157940321358616659221758088517262450417157546236820\\
&86392307057226733351528461327765206528690166061528491609355927358936857161005426154\\
&07200727390691671962634623969726562500000000000000000000000000000000000000000000000\\
&00\,x^3-1828221095436353717463174614437247330976158382656049872238375633162292868733\\
&74813398644462779979827289360372387537870206749333879123017694494738796239408112733\\
&28449200979037008209913975189192955946180380633322654605933402448500061893467638415\\
&60004136598159867314017545571506612004026822799245375407831045735371095310392708232\\
&81344476562979634640782750717162415439593036061033323032854282308841815501351114416\\
&935677549582015625000000000000000000000000\,x^4+3611743405663572269901520771342393081\\
&615670418917668079657497497275379573689559782204857676561201426730734937864394846343\\
&145339533029221076753457687358806578957577220325123249847686068331444194396137836564\\
&851492258428166002257460998695324872671642584396813257801785611761203007758151123590\\
&846908325491660994943573867336626008249075660058584580343705511632800783660690250144\\
&68919793348940488399910890856318533605156447765637099923292983683865980527\,x^5~.
\end{align*}

\begin{align*}
&\bullet r=2,~~k=8:\\
&-134359730992479741191539993021929240046707667243497260643608787066715685367224215\\
&737491611861837195833912916531295635207043101279109779464815792637735429160556081519\\
&781114488924354896071277498237553895626003602035163408030213775466890051884123979696\\
&892709068011402492330515480180433120829469951332425922338464907568322814768180251121\\
&52099609375-19360186351791045607549692954791951034992333280696911015169254722289053\\
&416284152526969102804765216978173499947700613776564480381020890372855544690651548265\\
&350440982743487680974528623114682454006214636514290830628924900453497304463562988472\\
&932081326581749238397927098301302152693352812062249144023151201693661661149265920000\\
&00000000000000000000000000000000000000000\,x-390316836654442678880830852748722871104\\
&342297141395464092338948279193236467065515866006754339081081230806552823618308925531\\
&062223061761394054417454859984429984530932997048108945642228333995340784233250345698\\
&236959509392506612879339512935989532924749377691029889790544558044773087305348204694\\
&968964402868306643966285330077585929060732569657251503013888000000000000000000000000\\
&000000\,x^2-1338957446769462324846303613619702797004750523035431458567179745765247566\\
&922671272624431292110245643326920094689892022102994901665786151216134233511274384197\\
&715005574803455795078830652501165515811832186248315634633782854587714488314168401954\\
&387419524934074760236295165494672383541915533427083824532101970644590493756309459429\\
&19986755048024228629588619715052355469949170735393164754944000000000000000\,x^3+162266\\
&7490347886753074861154430756407401826887961441735670196897840250173241848861777074876\\
&0912315207854109864386948761540569773284583982440580996741594147768641355272296624733\\
&2282632689165951791212346126162750936571776991055102255221295651635756525040838400078\\
&4931078753754990486512243476577338512702326541842225781394069607390183337578732342853\\
&96141178600267268748937574131999338755502611890176\,x^4~.
\end{align*}

\section*{Acknowledgements} 
We thank Tom Mainiero, Andrew Neitzke, Thorsten Weist, and Xi Yin for discussions.  The work of CC is support by a Junior Fellowship at the Harvard Society of Fellows.  The work of SHS is supported by the Kao Fellowship at Harvard University.

\appendix

\section{Tables of Wall-Crossing Data}
\label{sec:walldata}

In this appendix, we record the explicit wall-crossing data used to study the slope when $k=4$ (see Figure \ref{fig:wall}).  We record only $\log(\Omega)$ to four significant digits.  Complete, integral values of indices are available upon request.
\begin{table}[h!]

\begin{align*}
\begin{array}{|c|c|c|}
\hline
 ~M~ & ~N~ & \text{log $\Omega $} \\
\hline
 1 & 1 & 1.386 \\
 1 & 2 & 1.792 \\
 1 & 3 & 1.386 \\
 1 & 4 & 0 \\
 2 & 2 & 2.773 \\
 2 & 3 & 4.060 \\
 2 & 4 & 4.025 \\
 2 & 5 & 4.060 \\
 2 & 6 & 2.773 \\
 2 & 7 & 1.792 \\
 3 & 3 & 4.970 \\
 3 & 4 & 6.555 \\
 3 & 5 & 7.142 \\
 3 & 6 & 6.898 \\
 3 & 7 & 7.142 \\
 3 & 8 & 6.555 \\
 3 & 9 & 4.970 \\
 3 & 10 & 4.060 \\
 3 & 11 & 1.386 \\
 4 & 4 & 7.398 \\
 4 & 5 & 9.183 \\
 4 & 6 & 9.950 \\
 4 & 7 & 10.43 \\
 4 & 8 & 10.09 \\
 4 & 9 & 10.43 \\
 4 & 10 & 9.950 \\
 4 & 11 & 9.183 \\
 4 & 12 & 7.398 \\
 4 & 13 & 6.555 \\
 4 & 14 & 4.025 \\
 4 & 15 & 0 \\
 \hline
\end{array}
~~~~~
\begin{array}{|c|c|c|}
\hline
 ~M~ & ~N~ & \text{log $\Omega $} \\
\hline
 5 & 5 & 9.986 \\
 5 & 6 & 11.90 \\
 5 & 7 & 12.93 \\
 5 & 8 & 13.56 \\
 5 & 9 & 13.83 \\
 5 & 10 & 13.43 \\
 5 & 11 & 13.83 \\
 5 & 12 & 13.56 \\
 5 & 13 & 12.93 \\
 5 & 14 & 11.90 \\
 5 & 15 & 9.986 \\
 5 & 16 & 9.183 \\
 5 & 17 & 7.142 \\
 5 & 18 & 4.060 \\
 6 & 6 & 12.67 \\
 6 & 7 & 14.67 \\
 6 & 8 & 15.81 \\
 6 & 9 & 16.53 \\
 6 & 10 & 17.11 \\
 6 & 11 & 17.32 \\
 6 & 12 & 16.88 \\
 6 & 13 & 17.32 \\
 6 & 14 & 17.11 \\
 6 & 15 & 16.53 \\
 6 & 16 & 15.81 \\
 6 & 17 & 14.67 \\
 6 & 18 & 12.67 \\
 6 & 19 & 11.90 \\
 6 & 20 & 9.950 \\
 6 & 21 & 6.898 \\
 6 & 22 & 2.773 \\
 \hline
\end{array}
~~~~~
\begin{array}{|c|c|c|}
\hline
 ~M~ & ~N~ & \text{log $\Omega $} \\
\hline
 7 & 7 & 15.43 \\
 7 & 8 & 17.50 \\
 7 & 9 & 18.78 \\
 7 & 10 & 19.69 \\
 7 & 11 & 20.34 \\
 7 & 12 & 20.76 \\
 7 & 13 & 20.87 \\
 7 & 14 & 20.40 \\
 7 & 15 & 20.87 \\
 7 & 16 & 20.76 \\
 7 & 17 & 20.34 \\
 7 & 18 & 19.69 \\
 7 & 19 & 18.78 \\
 7 & 20 & 17.50 \\
 7 & 21 & 15.43 \\
 7 & 22 & 14.67 \\
 7 & 23 & 12.93 \\
 7 & 24 & 10.43 \\
 7 & 25 & 7.142 \\
 7 & 26 & 1.792 \\
 8 & 8 & 18.24 \\
 8 & 9 & 20.36 \\
 8 & 10 & 21.71 \\
 8 & 11 & 22.75 \\
 8 & 12 & 23.39 \\
 8 & 13 & 24.08 \\
 8 & 14 & 24.40 \\
 8 & 15 & 24.46 \\
 8 & 16 & 23.97 \\
 8 & 17 & 24.46 \\
 8 & 18 & 24.40 \\
 \hline
\end{array}
~~~~~
\begin{array}{|c|c|c|}
\hline
 ~M~ & ~N~ & \text{log $\Omega $} \\
\hline
 8 & 19 & 24.08 \\
 8 & 20 & 23.39 \\
 8 & 21 & 22.75 \\
 8 & 22 & 21.71 \\
 8 & 23 & 20.36 \\
 8 & 24 & 18.24 \\
 8 & 25 & 17.50 \\
 8 & 26 & 15.81 \\
 8 & 27 & 13.56 \\
 8 & 28 & 10.09 \\
 8 & 29 & 6.555 \\
 9 & 9 & 21.08 \\
 9 & 10 & 23.24 \\
 9 & 11 & 24.69 \\
 9 & 12 & 25.76 \\
 9 & 13 & 26.65 \\
 9 & 14 & 27.32 \\
 9 & 15 & 27.78 \\
 9 & 16 & 28.09 \\
 9 & 17 & 28.08 \\
 9 & 18 & 27.57 \\
 9 & 19 & 28.08 \\
 9 & 20 & 28.09 \\
 9 & 21 & 27.78 \\
 9 & 22 & 27.32 \\
 9 & 23 & 26.65 \\
 9 & 24 & 25.76 \\
 9 & 25 & 24.69 \\
 9 & 26 & 23.24 \\
 9 & 27 & 21.08 \\
 9 & 28 & 20.36 \\
 \hline
\end{array}
\end{align*}
\caption{The   wall-crossing data for $\log \Omega(M,N,k)$ with $k=4$ and $M+N\le 40$ . The plots for some of these data are shown in Figure \ref{fig:wall}.}
\end{table}

\begin{table}[h!]
\begin{align*}
\begin{array}{|c|c|c|}
\hline
 ~M~ & ~N~ & \text{log $\Omega $} \\
\hline
 9 & 29 & 18.78 \\
 9 & 30 & 16.53 \\
 9 & 31 & 13.83 \\
 10 & 10 & 23.96 \\
 10 & 11 & 26.15 \\
 10 & 12 & 27.66 \\
 10 & 13 & 28.85 \\
 10 & 14 & 29.80 \\
 10 & 15 & 30.41 \\
 10 & 16 & 31.14 \\
 10 & 17 & 31.56 \\
 10 & 18 & 31.77 \\
 10 & 19 & 31.74 \\
 10 & 20 & 31.21 \\
 10 & 21 & 31.74 \\
 10 & 22 & 31.77 \\
 10 & 23 & 31.56 \\
 10 & 24 & 31.14 \\
 10 & 25 & 30.41 \\
 10 & 26 & 29.80 \\
 10 & 27 & 28.85 \\
 10 & 28 & 27.66 \\
 10 & 29 & 26.15 \\
 10 & 30 & 23.96 \\
 11 & 11 & 26.86 \\
 11 & 12 & 29.08 \\
 11 & 13 & 30.65 \\
 11 & 14 & 31.91 \\
 11 & 15 & 32.93 \\
 11 & 16 & 33.74 \\
 11 & 17 & 34.41 \\
 \hline
\end{array}
~~~~~
\begin{array}{|c|c|c|}
\hline
 ~M~ & ~N~ & \text{log $\Omega $} \\
\hline
 11 & 18 & 34.96 \\
 11 & 19 & 35.32 \\
 11 & 20 & 35.49 \\
 11 & 21 & 35.41 \\
 11 & 22 & 34.87 \\
 11 & 23 & 35.41 \\
 11 & 24 & 35.49 \\
 11 & 25 & 35.32 \\
 11 & 26 & 34.96 \\
 11 & 27 & 34.41 \\
 11 & 28 & 33.74 \\
 11 & 29 & 32.93 \\
 12 & 12 & 29.78 \\
 12 & 13 & 32.03 \\
 12 & 14 & 33.64 \\
 12 & 15 & 34.94 \\
 12 & 16 & 36.00 \\
 12 & 17 & 36.95 \\
 12 & 18 & 37.54 \\
 12 & 19 & 38.31 \\
 12 & 20 & 38.73 \\
 12 & 21 & 39.07 \\
 12 & 22 & 39.21 \\
 12 & 23 & 39.10 \\
 12 & 24 & 38.56 \\
 12 & 25 & 39.10 \\
 12 & 26 & 39.21 \\
 12 & 27 & 39.07 \\
 12 & 28 & 38.73 \\
 13 & 13 & 32.72 \\
 13 & 14 & 34.99 \\
 \hline
\end{array}
~~~~~
\begin{array}{|c|c|c|}
\hline
 ~M~ & ~N~ & \text{log $\Omega $} \\
\hline
 13 & 15 & 36.65 \\
 13 & 16 & 38.01 \\
 13 & 17 & 39.16 \\
 13 & 18 & 40.13 \\
 13 & 19 & 40.90 \\
 13 & 20 & 41.58 \\
 13 & 21 & 42.18 \\
 13 & 22 & 42.59 \\
 13 & 23 & 42.86 \\
 13 & 24 & 42.95 \\
 13 & 25 & 42.81 \\
 13 & 26 & 42.26 \\
 13 & 27 & 42.81 \\
 14 & 14 & 35.68 \\
 14 & 15 & 37.96 \\
 14 & 16 & 39.65 \\
 14 & 17 & 41.07 \\
 14 & 18 & 42.27 \\
 14 & 19 & 43.29 \\
 14 & 20 & 44.16 \\
 14 & 21 & 44.73 \\
 14 & 22 & 45.53 \\
 14 & 23 & 46.04 \\
 14 & 24 & 46.41 \\
 14 & 25 & 46.64 \\
 14 & 26 & 46.69 \\
 15 & 15 & 38.65 \\
 15 & 16 & 40.94 \\
 15 & 17 & 42.67 \\
 15 & 18 & 44.11 \\
 15 & 19 & 45.37 \\
 \hline
\end{array}
~~~~~
\begin{array}{|c|c|c|}
\hline
 ~M~ & ~N~ & \text{log $\Omega $} \\
\hline
 15 & 20 & 46.39 \\
 15 & 21 & 47.37 \\
 15 & 22 & 48.13 \\
 15 & 23 & 48.81 \\
 15 & 24 & 49.43 \\
 15 & 25 & 49.85 \\
 16 & 16 & 41.62 \\
 16 & 17 & 43.93 \\
 16 & 18 & 45.69 \\
 16 & 19 & 47.19 \\
 16 & 20 & 48.45 \\
 16 & 21 & 49.59 \\
 16 & 22 & 50.58 \\
 16 & 23 & 51.42 \\
 16 & 24 & 51.97 \\
 17 & 17 & 44.61 \\
 17 & 18 & 46.93 \\
 17 & 19 & 48.72 \\
 17 & 20 & 50.25 \\
 17 & 21 & 51.58 \\
 17 & 22 & 52.75 \\
 17 & 23 & 53.76 \\
 18 & 18 & 47.61 \\
 18 & 19 & 49.94 \\
 18 & 20 & 51.75 \\
 18 & 21 & 53.30 \\
 18 & 22 & 54.68 \\
 19 & 19 & 50.61 \\
 19 & 20 & 52.95 \\
 19 & 21 & 54.79 \\
 20 & 20 & 53.63 \\
 \hline
\end{array}
\end{align*}
\caption{ The  wall-crossing data for $\log \Omega(M,N,k)$ with $k=4$ and $M+N\le 40$. The plots for some of these data are shown in Figure \ref{fig:wall}.}
\end{table}

%\bibliography{KroneckerQuiver}
\bibliographystyle{utphys}

\providecommand{\href}[2]{#2}\begingroup\raggedright\endgroup

\end{document}